\documentclass[english,aps,reprint]{revtex4-1}
\usepackage{graphicx}
\usepackage{subfigure}
\usepackage{hyperref}

\usepackage{physics}
\usepackage{floatrow}
\usepackage{amssymb}
\usepackage{xfrac}
\usepackage{amsmath}

\newcommand{\molpro}{\texttt{MOLPRO}}
\newcommand{\duo}{\texttt{DUO}}
\newcommand{\wn}{~cm\(^{-1}\)}
\newcommand{\Xst}{\text{X}\ensuremath{^2\Sigma^+}}
\newcommand{\Hst}{\text{H}\ensuremath{^2\Delta}}
\newcommand{\Ast}{\text{A}\ensuremath{^2\Pi}}
\newcommand{\Bst}{\text{B}\ensuremath{^2\Sigma^+}}

\newcommand{\bahiso}[2]{\ensuremath{^{#1}\text{Ba}^{#2}\text{H}}}

\begin{document}

\title{Assignment of excited-state bond lengths using branching-ratio measurements: \\ The B$^2\Sigma^+$ state of BaH molecules}

\author{K. Moore}
\affiliation{School of Chemistry and Chemical Engineering, Queen's University Belfast, Stranmillis Road, Belfast BT9 5AG, Northern Ireland, UK}

\author{R. L. McNally}
\affiliation{Department of Physics, Columbia University, 538 West 120th Street, New York, NY 10027-5255, USA}

\author{T. Zelevinsky}
\email{tanya.zelevinsky@columbia.edu}
\affiliation{Department of Physics, Columbia University, 538 West 120th Street, New York, NY 10027-5255, USA}

\author{I. C. Lane}
\email{i.lane@qub.ac.uk}
\affiliation{School of Chemistry and Chemical Engineering, Queen's University Belfast, Stranmillis Road, Belfast BT9 5AG, Northern Ireland, UK}

\begin{abstract}
Vibrational branching ratios in the B$^2\Sigma^+$ -- X$^2\Sigma^+$ and A$^2\Pi$ -- X$^2\Sigma^+$  optical-cycling transitions of BaH molecules are investigated using measurements and {\it ab initio} calculations. The experimental values are determined using fluorescence and absorption detection.  The observed branching ratios have a very sensitive dependence on the difference in the equilibrium bond length between the excited and ground state, $\Delta r_e$:  a 1 pm (.5\%) displacement can have a 25\% effect on the branching ratios but only a 1\% effect on the lifetime.  The measurements are combined with theoretical calculations to reveal a preference for a particular set of published spectroscopic values for the B$^2\Sigma^+$ state ($\Delta r_e^{B-X}$ = +5.733 pm), while a larger bond-length difference ($\Delta r_e^{B-X} = 6.3-6.7$ pm) would match the branching-ratio data even better.  By contrast, the observed branching ratio for the  A$^2\Pi_{3/2}$ -- X$^2\Sigma^+$ transition is in excellent agreement with both the {\it ab initio} result and the spectroscopically measured bond lengths.  This shows that care must be taken when estimating branching ratios for molecular laser cooling candidates, as small errors in bond-length measurements can have outsize effects on the suitability for laser cooling.  Additionally, our calculations agree more closely with experimental values of the B$^2\Sigma^+$ state lifetime and spin-rotation constant, and revise the predicted lifetime of the H$^2\Delta$ state
to 9.5 $\mu$s.
\end{abstract}

\maketitle

\section{Introduction}

Spectroscopy is one of the most precise measurement tools in physical chemistry.  A typical parameter determined using such techniques is $r_e$, the equilibrium bond length that is often reported with an uncertainty of 1 fm or $<10^{-5}$, less than the width of an atomic nucleus.  By contrast, a bond length calculated using quantum chemistry methods within 1 pm of the experimental value is regarded as very good, especially for excited states, and one within 0.1 pm is regarded as state of the art except for very small molecules.  Even BeH, with only five electrons, presents challenges for theory \cite{Dattani2015}.
{\it Ab initio} quantum chemistry, however, directly calculates values such as $r_e$ that spectroscopic studies only infer via the determination of $B_e$, the equilibrium rotational constant.  Furthermore, $B_e$ itself cannot be directly measured: instead, it is calculated from the measured rotational constants $B_v$ for at least two vibrational levels $v$. Spectroscopic methods are very reliable when a single isolated potential-energy curve is under analysis but become less robust when several potential curves are closely spaced in energy and interact strongly. In such cases a model must be applied to the coupled potentials, and its details influence the derived values of constants such as $B_e$.

In this paper, a combination of branching-ratio measurements and calculations is used to distinguish between spectroscopic determinations of $r_e$ that were originally reported to five decimal places but disagree at the second decimal place.  Furthermore, we report a bond length that is consistent with our branching-ratio measurements.
The sensitivity of our method relies on monitoring the branching ratios of highly diagonal ($\Delta v=0$) transitions, where a small change in the decay to additional quantum states results in a large increase in the relative populations of those states.
We find that the branching ratios sensitively depend on the difference between the excited- and ground-state equilibrium bond lengths.
An example of a molecule with this property is barium monohydride, BaH, a radical of interest for direct laser cooling \cite{Lane2015, Tarallo2016, Iwata2017}.  For BaH, a 1-pm (0.5\%) relative bond-length displacement can have a 25\% effect on the branching ratios but only a 1\% effect on the natural lifetimes.  Conversely, this sensitivity implies that care must be taken when theoretically evaluating the laser cooling prospects of new molecular candidates, since branching ratios $-$ key parameters for laser cooling $-$ are strongly affected by small errors in the relative bond length of the electronic states used in the cooling scheme.

\section{Spectroscopy background}

The optical and near-infrared spectra of BaH \cite{Watson1933, Watson1935, Koontz1935, Kopp1966-1, Kopp1966-2, Huber1979, Appelblad1985, Fabre1987, Bernard1987, Magg1988, Bernard1989, Barrow1991, Walker1993, Berg1997, Ram2013} are dominated by the three 5$d$-complex states that correlate to the $5d$ state of the Ba atom:  B$^2\Sigma^+$, A$^2\Pi$, and H$^2\Delta$.  Since all three reach below the ground-state dissociation threshold, the only decay mechanisms are radiative.  In addition, these three low-lying excited states possess spectroscopic parameters that closely resemble the X$^2\Sigma^+$ ground state.
The resulting diagonal Franck-Condon (FC) factors ensure absorption-emission cycles numbering in the thousands, as required for efficient laser cooling with as few optical fields as possible. A buffer gas beam \cite{Hutzler2012} of BaH molecules in the  X$^2\Sigma^+_{1/2}$ $v^{\prime\prime}$~=~0, $N^{\prime\prime}$~=~1 state was recently demonstrated \cite{Iwata2017} and used in precise measurements of the ($0-0$) and ($0-1$) branching ratios, where ($v^{\prime} - v^{\prime\prime}$) denotes an electronic transition between the lower $v^{\prime\prime}$ vibrational level of X$^2\Sigma^+$ and the $v^{\prime}$ level in the excited state B$^2\Sigma^+ $.  In addition, the quantum state purity of the buffer gas beam was exploited in measurements of the magnetic $g$ factors and hyperfine structure of the lowest rovibronic levels of the A$^2\Pi_{1/2}$ and B$^2\Sigma^+_{1/2}$ states \cite{Iwata2017} in preparation for laser cooling experiments.

One of the potential laser cooling transitions, B$^2\Sigma^+ $ -- X$^2\Sigma^+$, is the focus of our paper.  The ($0-0$) and ($1-1$) vibronic bands of BaH were first reported in 1933 \cite{Watson1933}, and soon extended \cite{Koontz1935} to include the much weaker off-diagonal bands ($1-0$) and ($2-1$). The equivalent study on BaD was published 30 years later \cite{Kopp1966-2}.  High-quality Fourier-transform data of BaH \cite{Appelblad1985} provided a hundred-fold improvement
in the accuracy of the reported spectroscopic constants and extended the analysis to the B$^2\Sigma^+$ $v^{\prime} =$ 0-3 vibronic levels, while the B$^2\Sigma^+$ $v^{\prime}$ = 0, $J$ = 11/2 level lifetime \cite{Berg1997} was measured to be 124(2) ns, where $J$ is the total angular momentum.

A more comprehensive analysis \cite{Bernard1989} attempted a simultaneous fit of spectroscopic data involving all the 5$d$-complex states,
where 1478 BaH spectral lines and 2101 BaD lines were used.
This is currently the only experimental measurement of the spin-orbit splitting in the \Hst\ state, $A$ = 217.298~cm$^{-1}$ for $v^{\prime}$ = 0.  There is, however, a disagreement between the  measured value of the spin-orbit separation in the A$^2\Pi$ state, $A$~=~341.2~cm$^{-1}$ for $v=0$, with an earlier value \cite{Kopp1966-2} of 483~cm$^{-1}$.  In addition, the spin-rotation constants for both the $v=0$ and $1$ levels in the B$^2\Sigma^+_{1/2}$ state are an order of magnitude smaller and of opposite sign to those from previous work \cite{Watson1933,Koontz1935,Appelblad1985}. Finally, due to the challenges of working with higher vibrational levels, the spectroscopic constants are limited \cite{Bernard1987,Bernard1989} to the $v=0$ and $1$ levels for all three $5d$ states.

Detailed information on higher-lying vibrational levels is missing for all the states, as diagonal transitions mean that only a small part of the potential curves can be explored spectroscopically.  In the absence of such experimental data, quantum chemistry can be used to explore these dark regions of the potentials \cite{Allouche1992, Moore2016, Moore2018}. {\it Ab initio} techniques have also been applied to understand the laser cooling process \cite{Wells2011, Lane2015, Gao2014}. The latest theoretical study on BaH (Ref. \cite{Moore2018}, here referred to as Moore18) includes spin-orbit coupling as well as other relativistic effects and a thorough analysis of multiple decay pathways. The computed spectroscopic constants for all three $5d$-complex excited states were in good agreement with experiments. However, in an effort to improve the reliability of the excited-state decay properties, the calculated potentials were shifted to within 0.1 pm of the experimental values for $r_e$. For the B$^2\Sigma^+$ state the chosen experimental data came from Appelblad {\it et al} \cite{Appelblad1985}.  This study was broadly consistent with earlier spectroscopic measurements \cite{Watson1933, Koontz1935} but had a greater reported precision.  Moreover, it broadened the published potential-energy and rotational constant ($T_v$ and $B_v$) data to cover the vibrational levels $v$ = 0 - 3, and it quoted $B_e$ (in cm$^{-1}$) to an accuracy of six decimal places.
Oddly, applying this correction reduced the agreement with the measured lifetime \cite{Berg1997} of the B$^2\Sigma^+$ state, although the difference was $<1.5\%$.  Also disappointingly, the agreement with the B$^2\Sigma^+_{1/2} -$ X$^2\Sigma^+_{1/2}$ branching-ratio measurement \cite{Tarallo2016} became significantly worse.

To shed light on these discrepancies, here we report an accurate measurement of the branching ratios on a cryogenic beam of BaH.  We combine the measurement with the most detailed {\it ab initio} work to date to confirm the correct bond length in the excited  B$^2\Sigma^+$ state. Adopting a slightly different approach to Moore18 is shown to improve both the spectroscopic constants and the calculated lifetime of the B$^2\Sigma^+$ state.
Adjusting the upper-state $r_e$ to match the work of Bernard {\it et al} \cite{Bernard1989} does significantly improve the agreement, although the theoretical value still remains outside the error bars of the experimental result.  Based on our measurements and quantum chemistry calculations, we suggest a value of $r_e$ that is based on very sensitive branching-ratio measurements rather than on spectroscopy alone.

\begin{figure}[!ht]
\includegraphics[width=85mm]{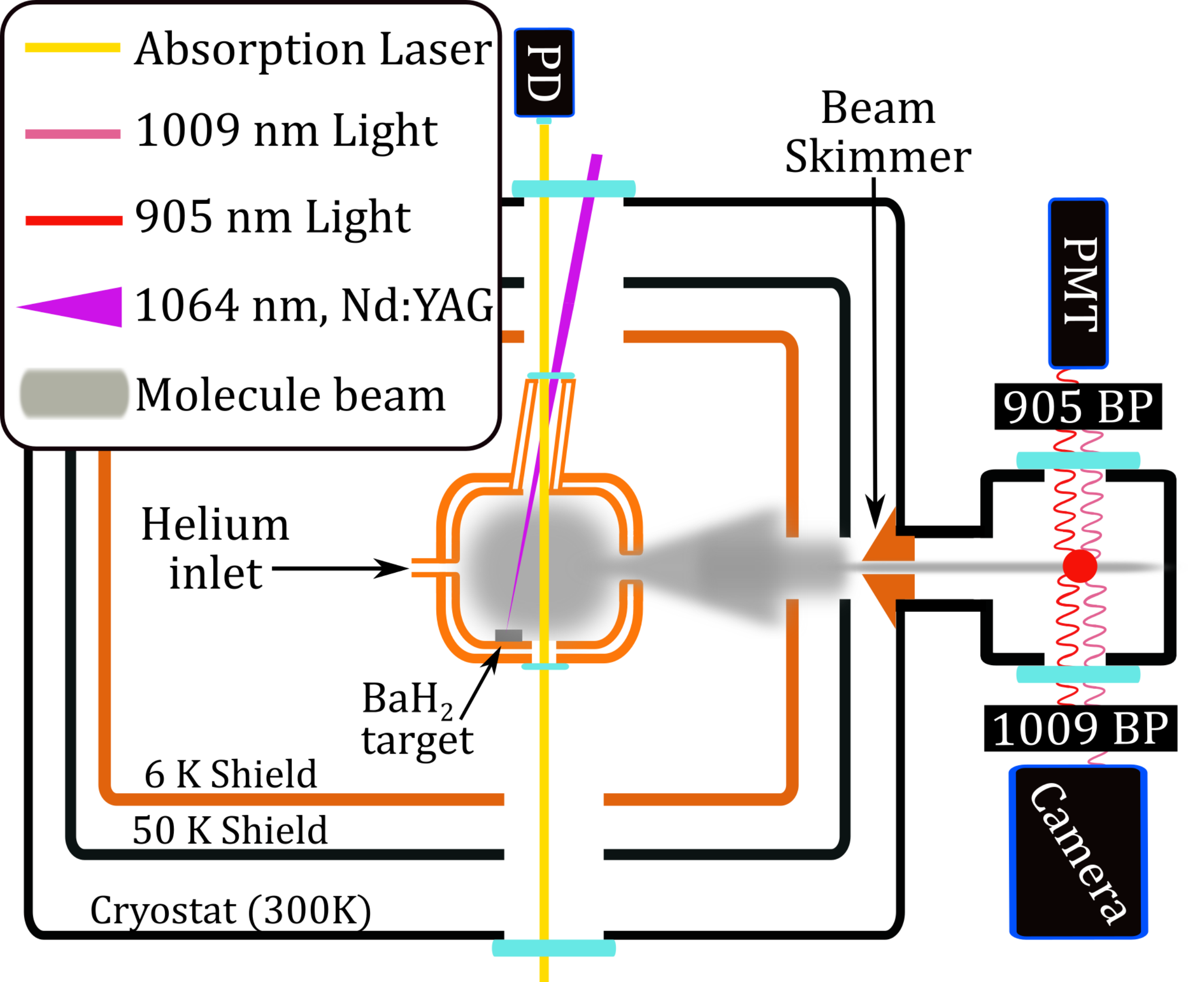}
\caption{Diagram of the buffer-gas-cooled molecular-beam source used in the experiment.  BaH molecules are generated via ablation inside a helium-filled copper cell, then swept through a series of apertures and beam skimmers to form a collimated beam.  The beam enters a detection region 40 cm downstream where resonant laser light (905 nm, represented by the red dot, with the beam propagating perpendicular to both the molecular beam and detectors) excites the molecules to the B$^2\Sigma^+$ $v=0$, $J=1/2$ level with (+) parity.  Two detectors monitor fluorescence to the X$^2\Sigma^+$ $v=0$ state at 905 nm and to the X$^2\Sigma^+$ $v=1$ state at 1009 nm.  Dichroic BP filters isolate specific decay channels for each detector.
The PMT is used to continuously monitor the beam flux.} \label{fig:VBR_measure}
\end{figure}

\section{Franck-Condon Factor ratio measurements}\label{experiment}
To perform measurements of the branching ratios for the relevant electronic transitions in BaH, we utilize two complementary techniques, both using our cryogenic BaH molecular beam \cite{Iwata2017}.  The schematic of the experiment is shown in Fig. \ref{fig:VBR_measure}.  The BaH beam is generated via ablation of a BaH$_2$ rock inside a helium-filled copper cell cooled to 6 K.  A pair of cell windows allow optical access for a resonant laser beam, which can be used to monitor absorption by measuring the transmitted power on a photodiode (PD).  The molecules rapidly thermalize with the helium and are swept out of the cell to form a beam.  The cold molecules then travel 40 cm downstream where they enter a detection region equipped with two types of photodetector:  a near-infrared-enhanced
photomultiplier tube (PMT) and a deep-depletion charge-coupled device (CCD) camera.
Each of the two detectors includes dichroic bandpass (BP) filters as shown in Fig. \ref{fig:VBR_measure} and is used for fluorescence detection.

Both absorption and emission measurements were made on the beam. The absorption technique is based on a differential measurement of the absorption cross section from one ground vibrational state $v''$ to two different excited vibrational levels $v'$ \cite{Kozyryev2015}.  We write the integrated absorption cross-section for a rovibronic transition \cite{Hilborn1982,Hansson2005} as
\begin{align*}
\sigma_{v'J' v'' J''}(\tilde{\nu})
&= \frac{g(c\tilde{\nu}) (2J'+1)}{ 8\pi(2J''+1)}\frac{A_{v'J',v''J''}}{(\tilde{\nu}_{v'J' v''J''}^{\vphantom{q}})^2}
\end{align*}
\begin{align*}
&= \frac{ 2 {\pi}^2 \tilde{\nu}_{v'J' v''J''} ^{\vphantom{q}}}{3 \varepsilon_{0} h (2J''+1)^{\vphantom{Q}}}g(c\tilde{\nu})S_{v'J',v''J''},
\end{align*}
where $ A_{v'J',v''J''}^{\vphantom{q}}$ is the Einstein $A$ coefficient describing the absorption to the $v' J'$ excited rovibrational state from the $v'' J''$ ground level, $\tilde{\nu}_{v'J' v''J''}^{\vphantom{q}}$ is the transition wave number, $\varepsilon_{0}$ is the vacuum permittivity, and $g(c\tilde{\nu})$ is the line-shape function.  The transition line strength is
\begin{align} \label{eqn:linestrength}
S_{v'J',v''J''}^{\vphantom{q}} &= \vert { \mel{ \Lambda ' v' J' \epsilon '}{\mu }{\Lambda '' v'' J'' \epsilon ''}} \vert ^2
&= | M |^{2} S_{J',J''}^{\vphantom{q}},
\end{align}
where $\epsilon$ is the parity label, $| M |^{2}$ is the square of the vibronic transition moment, and $S_{J',J''}^{\vphantom{q}}$ is the Honl-London factor for the transition \cite{Watson2008}.
As all the transitions studied originate on a level where $J=1/2$, and the principal results are the ratios between transitions with identical $\Delta J = J' - J''$ values, we drop the explicit references to $J$.   The $| M |^{2}$ is approximately the product of $q_{v' v''}^{\vphantom{q}}$, the transition's FC factor, and $\vert R_e \vert ^2$, the square of the electronic transition dipole moment (all the transitions studied here preserve the electronic spin),
\begin{equation*}
| M |^{2} = \vert { \mel{ \Lambda ' v'}{\mu }{\Lambda '' v''}} \vert ^2 \approx q_{v' v''}^{\vphantom{q}}\abs{R_e}^2.
\end{equation*}
In the results we quote the FC factors $ q_{v' v''}^{\vphantom{q}}$, although we technically measure the  vibronic moments.

We relate the cross section $\sigma_{v' v''}^{\vphantom{q}} (\omega)$ to an experimentally measurable optical absorption  to give the absorbance  $\mbox{\large \texttt{A}}(v' , v'')$:
\begin{equation*}
\frac{\Delta I}{I}= 1 - e^{-N \sigma_{v' v''}(\omega) l} \equiv 1-e^{-\mbox{\large \texttt{A}}(v' , v'')},
\end{equation*}
where $N$ is the molecular density, $\omega$ is the angular frequency of light, and $l$ is the path length.  By taking the ratio of absorption between two transitions that differ only in the excited-state vibrational number ($v'=0$ or 1), we can cancel the dependence on every parameter except $ q_{v' v''}$ and $\tilde{\nu}$, where the latter is known to a high accuracy.  We can then relate the experimentally measured absorption ratio to the ratio of the FC factors for the transition pairs, or the absorption vibronic transition ratio:
\begin{equation}
\mathrm{VTR} = \frac{\mbox{\large \texttt{A}}(v_1 , v'') \tilde{\nu}_{v_2 v''}}{\vphantom{\tilde{A}}\mbox{\large \texttt{A}}^{}(v_2 = v'', v'') \tilde{\nu}_{v_1 v''}^{}} \approx \frac{q_{\vphantom{q} v_1 v''}^{}}{\vphantom{\tilde{A}} q_{v_2 v''}^{}}.
\label{eq:VTR-A}
\end{equation}
This definition ensures that the quoted VTR is always $<1$ for a diagonal system such as the electronic transitions in BaH.  Measurements of the VTR using this technique are consequently invariant to fluctuations of the molecular density and rely only on quantities we can accurately determine.

The measurement of the emission VTR relies on a direct observation of a decay probability ratio.  The ratio $\mathcal{R}_{v'v''}$ of the measured emission to the total overall decay rate (branching ratio) can be expressed in terms of the transition FC factor \cite{Barry2013}:
\begin{equation}  \label{eq:exp_VBR}
\mathcal{R}_{v' v''} = \frac{q_{\vphantom{q} v' v''}^{} \tilde{\nu}_{v' v''}^3}{\sum\limits_{k=0}^{\infty} q_{v',k}^{} \tilde{\nu}_{v',k}^3},
\end{equation}
where the summation is over all available radiative decay channels.
By observing simultaneous fluorescence from a single excited rovibrational state $v'=0$, $J' = 1/2$ to two different vibrational ground states, $v''=0$ and 1, we directly compare the relative decays $\mathcal{R}_{0 v''}$ and determine the $ q_{v' v''}^{}$ ratio for the two transitions:
\begin{equation}  \label{eq:exp_VTR}
\mathrm{VTR} = \frac{I_{0v_1}^{} \tilde{\nu}_{0v_2}^3}{I_{0v_2}^{} \tilde{\nu}_{0v_1}^3} \approx \frac{q_{0v_1}^{}}{ q_{0 v_2}^{}},
\end{equation}
where $I_{0v_i}$ in the intensity of the observed decay to the $i$th ground vibrational state.  These complementary techniques allow us to measure a variety of $ q_{v' v''}^{}$ ratios using a series of differential measurements.  The experimental results, and the comparison to theoretical work presented below, are provided in Table \ref{tab:bah_comparisons}.  All the ground-state rovibrational levels involved are of ($-$) parity ($N''=1$).

\begin{table}
\centering
\caption{Comparison between the present experimental measurements of the vibronic transition ratios (VTRs) and the corrected theoretical results.
The ground state is the Morse/long-range (MLR) potential based on ACV$n$Z/CBS MLR calculations \cite{Moore2016} and the excited states are based on ACV$Q$Z potentials. The difference in equilibrium bond lengths between the excited and ground states, $\Delta r_e$, is set at the value proposed by Bernard {\it et al} \cite{Bernard1989} for the theoretical values, and the last column corresponds to the longer excited-state bond length proposed here for B$^2\Sigma^+$.}

\label{tab:bah_comparisons}
{\footnotesize
\resizebox{\columnwidth}{!}{\begin{tabular}{|c|c|c|c|c|}
\noalign{\vskip 4mm}
\hline
\rule{0pt}{2.5ex}Electronic	& VTR$^a$	& Experimental	&Theoretical$^b$ & Proposed$^c$ \\
			transitions	&			& value		&	&  	\\
\hline
\rule{0pt}{3.5ex}A$^2\Pi_{3/2}$ $\leftarrow$ X$^2\Sigma^+$	
	&	\( \sfrac{\mbox{\small \it q}_{\mbox{\tiny 10}}}{\mbox{\small \it q}_{\mbox{\tiny 00}}} \)	&	0.037(2)	&0.037 &0.037
\rule{0pt}{2.5ex}	\\
B$^2\Sigma^+$ $\leftarrow$ X$^2\Sigma^+$
	&	\( \sfrac{\mbox{\small \it q}_{\mbox{\tiny 10}}}{\mbox{\small \it q}_{\mbox{\tiny 00}}} \)  &	0.072(6)	&0.059&0.076
\rule{0pt}{2.5ex}	\\
B$^2\Sigma^+$ $\leftarrow$ X$^2\Sigma^+$	
	&	\( \sfrac{\mbox{\small \it q}_{\mbox{\tiny 01}}}{\mbox{\small \it q}_{\mbox{\tiny 11}}} \)	&	0.115(5)	&0.072&0.118
\rule{0pt}{2.5ex}	\\
B$^2\Sigma^+$ $\rightarrow$ X$^2\Sigma^+$	
	&	\( \sfrac{\mbox{\small \it q}_{\mbox{\tiny 01}}}{\mbox{\small \it q}_{\mbox{\tiny 00}}} \) 	&	0.092(20)	&0.045& 0.065
\rule{0pt}{2.5ex}	\\
\hline
\noalign{\vskip 1.8mm}

\multicolumn{4}{l}{\textsuperscript{a} $J'' = J'=1/2$ for the absorption lines.}\\
\multicolumn{4}{l}{\textsuperscript{b} Bernard {\it et al} \cite{Bernard1989} $r_e$ value used}\\
\multicolumn{4}{l}{\textsuperscript{c} With the proposed +1.5 pm shift in the B$^2\Sigma^+$ bond length}\\
\end{tabular}}}
\end{table}

\subsection{Relative absorption measurements}
For absorption measurements we utilize the high molecular density inside the cryogenic cell (Fig. \ref{fig:VBR_measure}).  We alternate the probe beam between two coaligned lasers, each tuned to the resonant frequency of the  energy  levels of interest.  This allows for real-time cancellation of any variability in the molecular yield, as the lasers intersect the same region of the cell.
The laser intensities are well below saturation, rendering the measurement insensitive to optical pumping.
This was confirmed by varying the absorption laser power over an order of magnitude, still below saturation, and observing no detectable difference in the resulting VTR.

\begin{figure}[!ht]
\includegraphics[width=85mm]{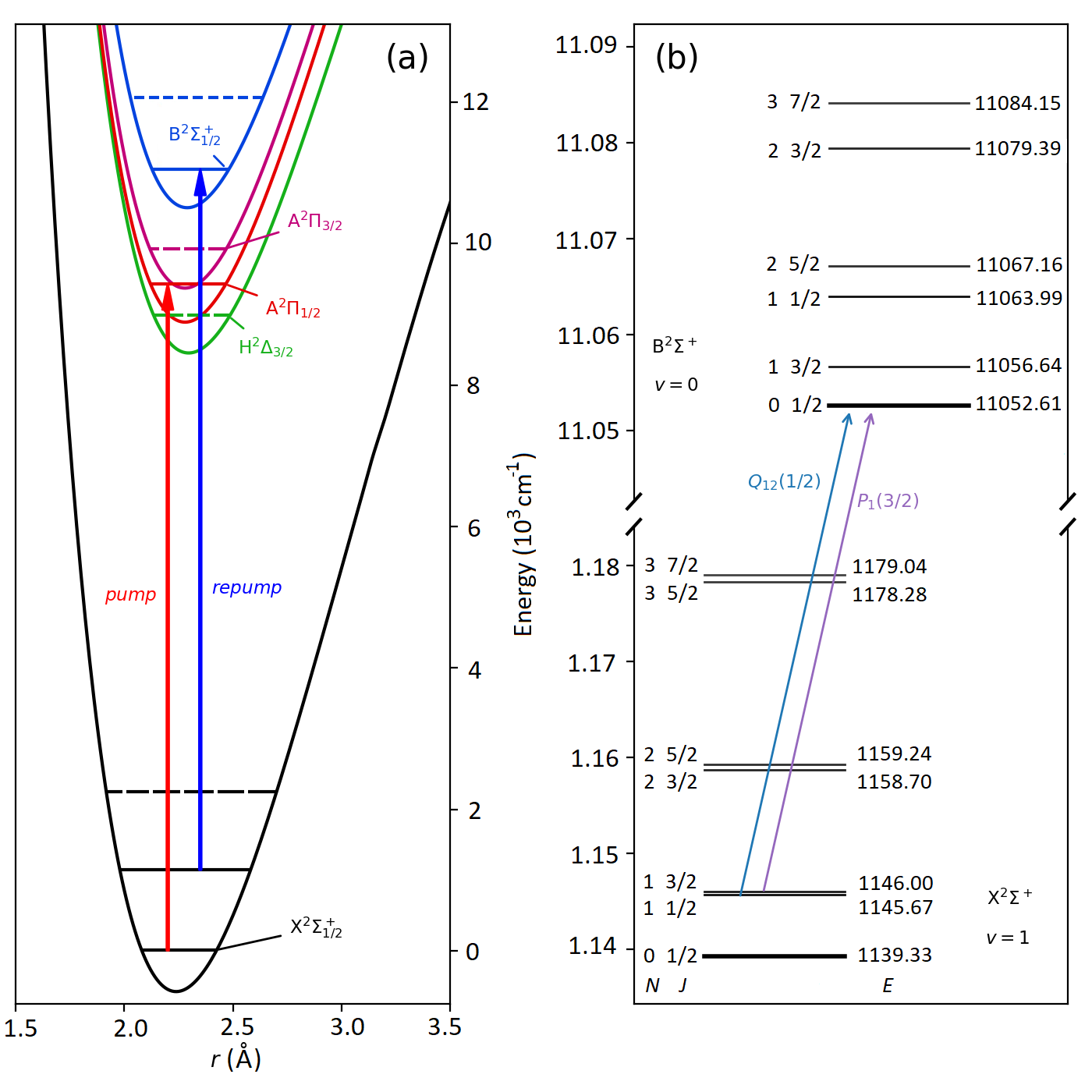}
\caption{(a) {\it Ab initio} potential-energy curves of the lowest electronic states involved in laser cooling BaH. The potential-energy curves and spin-orbit matrix elements are determined by the MRCI method (with Davidson correction) using \molpro\  \cite{Werner2010} with the ACV$Q$Z basis set for Ba and a similar basis set for H.
The H$^2\Delta$ and A$^2\Pi$ potentials lie below B$^2\Sigma^+$ in the FC region.  The vertical arrows correspond to the main cooling and repumping transitions.  The vibrational levels directly involved in cooling are marked by solid lines, while the dashed lines correspond to the principal losses (unpumped vibrational levels), specifically the H$^2\Delta_{3/2}$ $v=0$ (largest single decay, green), A$^2\Pi_{3/2}$ $v=0$ (purple) and X$^2\Sigma_{1/2}^+$ $v=2$ (black) levels.  Also marked is the B\(^2\Sigma_{1/2}^+, v=1 \) level (dotted blue line) involved in the absorption measurements reported here.  (b) Calculated BaH rotational structure in the lowest $^2\Sigma^+$ states. The ACV$Q$Z X$^2\Sigma^+$ was replaced by a more accurate MLR potential based on a complete basis set (CBS) calculation \cite{Moore2016}.
The lowest rotational energies for each vibronic level are shown.
The vibrational spacings correspond to the {\it ab initio} values determined by \duo\ with the lowest vibrational level in each state anchored to experimental measurements.
The zero-energy reference is the lowest rovibrational level of X$^2\Sigma^+$, 580 cm$^{-1}$ above the potential minimum. Also marked is the B\(^2\Sigma_{1/2}^+\; v=0, N=0 \leftarrow \)  X\(^2\Sigma_{1/2}^+\; v=1, N=1\) repumping transition proposed for laser cooling.} \label{spin_rotation}
\end{figure}

\begin{figure}[!ht]
\includegraphics[width=90mm]{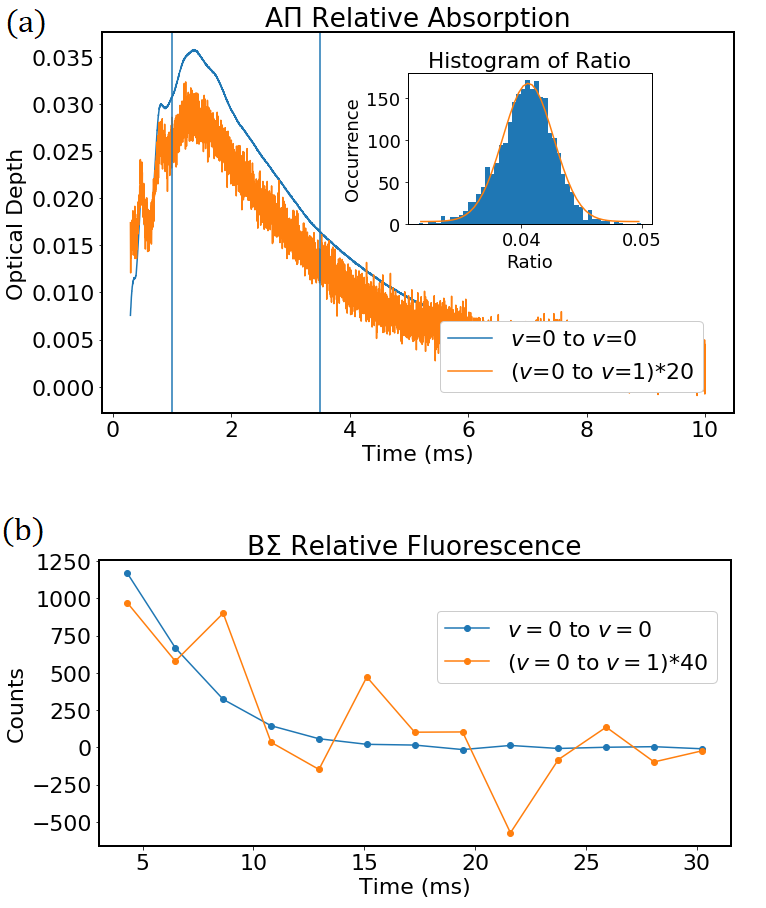}
\caption{(a) Relative absorption on the $(1-0)$ and $(0-0)$ A$^2\Pi_{3/2}$ electronic transitions, showing the average absorption on each transition over 200 experimental shots.  Vertical bars indicate the region of the signal used to calculate the absorption ratio, and the inset histogram shows the distribution of instantaneous ratios over this region.  Error bars for the absorption ratio are conservatively taken as the FWHM of the distribution.  This absorption ratio can be related to the FC factor ratio using Eq. (\ref{eq:VTR-A}).
(b) Relative fluorescence signal from the B$^2\Sigma^+ v=0$ $\rightarrow$ X$^2\Sigma^+ v=1$
and  $0$ decay paths.  Each point corresponds to the summed counts on the CCD image taken at a given time after ablation with background counts removed.  These traces are averaged over 100 shots.  Early times were removed since ablation light corrupts the CCD images.  Negative counts arise from noise in the background subtraction.  The small signal in the X$^2\Sigma^+ v=1$ decay path dominates the uncertainty in this measurement.
} \label{fig:Abs_Ratio}
\end{figure}

We performed three measurements, obtaining $q_{10}/q_{00}$ for the B$^2\Sigma^+$ $\leftarrow$ X$^2\Sigma^+$ and A$^2\Pi_{3/2}$ $\leftarrow$ X$^2\Sigma^+$ electronic transitions, and $q_{01}/q_{11}$ for B$^2\Sigma^+$ $\leftarrow$ X$^2\Sigma^+$.  To measure $q_{01}/q_{11}$, an additional laser was coaligned with the absorption lasers and tuned to the  A$^2\Pi_{3/2} v=1$ $\leftarrow$ X$^2\Sigma^+$ $v=0$ transition. This is required to increase the population in the X$^2\Sigma^+$ $v''=1$ state, as the $v''=1$ population is negligibly small for BaH thermalized to 6 K \cite{Tarallo2016}.  Only the $Q_{12}$ rotational lines shown in Fig. \ref{spin_rotation} were measured for each transition.  Figure \ref{fig:Abs_Ratio}(a) shows the absorption signals and their ratio.
For very short times below 1 ms, we observe varying ratios.  However, after the early dynamics dissipate, the ratios remain constant between 1 and 10 ms.  As shown, we select the 1-3.5 ms time window, since at longer times the ratio becomes dominated by noise.
The experimental values for the FC factor ratios thus obtained with the use of Eq. (\ref{eq:VTR-A}) are presented in Table \ref{tab:bah_comparisons} (top three entries).

\subsection{Direct fluorescence detection}

To perform a direct measurement of the emission VTR in the decay of the B$^2\Sigma^+$ $v'=0$ state, near-infrared fluorescence is recorded from the molecular beam.  The measurement is carried out via simultaneous collection of spontaneous emission light as the molecules decay from B$^2\Sigma^+$ $v'=0$ to X$^2\Sigma^+$ $v''=0$ at 905.3 nm and to X$^2\Sigma^+$ $v''=1$ at 1009.4 nm.  The emission VTR is then related to the FC factor ratio as in Eq. (\ref{eq:exp_VTR}).

In the detection region, the molecules are excited to the B$^2\Sigma^+$ $v'=0$, $J=1/2$ (+) level using two external-cavity diode lasers (linewidth $\sim1$ MHz) on resonance with the
\begin{equation*} B^2\Sigma^+ (v=0, N=0, J=1/2) \leftarrow
X^2\Sigma^+ (0, 1, 1/2)
\end{equation*}
and
\begin{equation*}
B^2\Sigma^+ (v=0, N=0, J=1/2) \leftarrow X^2\Sigma^+ (0, 1, 3/2)
\end{equation*}
transitions, forming a quasi-closed system.    Prior to the VTR measurement, calibration of the relative efficiency of each collection system is performed by placing a 905-nm BP filter in front of each detector (Fig. \ref{fig:VBR_measure}) and collecting the fluorescence on both.  A 1009-nm BP filter was then placed at the CCD camera and simultaneous measurements of fluorescence to the X$^2\Sigma^+$ $v''=0$ and $1$ states were taken with the PMT and CCD camera, respectively.  The ratio of the signals seen on the two detectors and the known efficiency of each filter and detector allow a measurement of the VTR that is independent of the molecule number.

The transmission of the bandpass filters was experimentally determined using a collimated laser and a power meter.  The effect of off-angle transmission through each filter was not found to be a significant factor, mainly due to the relatively large bandwidths of the filters.  The CCD had a particularly high sensitivity in the near-infrared (70\% at 1009 nm and 96\% at 905 nm).  Factory calibration was used to estimate the efficiency, since the fluorescence measurement can tolerate a substantial error in the calibration due to its relatively large statistical uncertainty.

Both $Q_{12}$ and $P_{1}$ rotational lines (Fig. \ref{spin_rotation}) were measured for each transition as they cannot be distinguished by the filters. The measured value of the $q_{01}/q_{00}$ VTR from data in Fig. \ref{fig:Abs_Ratio}(b) is 0.092(20).  The quoted statistical uncertainty is dominated by shot noise due to the relatively poor ($\sim$ 1\%) quantum efficiency of the PMT at 905 nm.

\section{Theoretical branching ratios}

Our theoretical quantum chemistry work concentrates on estimating three observables that present rigorous tests for different aspects of the {\it ab  initio} calculations:
\begin{enumerate}
\item B$^2\Sigma^+$  state spin-rotation constants:  the spin-orbit and ladder matrix elements;
\item B$^2\Sigma^+$  state lifetimes:  transition dipole moments;
\item B$^2\Sigma^+$ $\rightarrow$ X$^2\Sigma^+$ branching ratios:  bond lengths and potential-energy functions.
\end{enumerate}
The present experimental study provides the required test for the final property, while previously published works \cite{Appelblad1985,Berg1997} provide the benchmarks for the first two.

\begin{table*}
\caption{Spectroscopic $T_e$ and $r_e$ values for the lowest electronic states of BaH. These have been determined from tabulated rotational constants $B_e$ or, if unavailable, by extrapolating $B_v$ values as outlined in the text. Theoretical values from Moore18 \cite{Moore2018} are also included.  Values selected as reference in Moore18  \cite{Moore2018} are in bold. Values in italics correspond to minor isotope data.
\(T_e\) values are relative to the corresponding \( T_e(\text{X}^2\Sigma^+)\)= 0~\wn, with the zero-point energy listed in parentheses for the X state where available.}
\label{exp_ref}
\centering\footnotesize
\begin{tabular}{|lllllll|}
\noalign{\vskip 1.6mm}
\hline
\rule{0pt}{3ex}State &
	\multicolumn{1}{c}{Isotope}	&
	\multicolumn{1}{c}{\(B_e\) / \wn} &
	\multicolumn{1}{c}{\(r_e\) / \AA} &
	\multicolumn{1}{c}{\(T_e\) / \wn} &
	Year &
	\multicolumn{1}{l|}{Reference}
	\rule{0pt}{2ex}\\
\hline
\rule{0pt}{4ex}X\(^2\Sigma^+\) &
	\bahiso{138}{1}	&
	3.359\(^{a(3)}\)	&
	2.239\(^d\)	&
	(573.35)	&
	2018	&
	Moore and Lane \cite{Moore2018}, theoretical \rule{0pt}{2.5ex}
	\\[1ex]	
&%
	\bahiso{138}{1}	&
	\textbf{3.38243550}	&
	\textbf{2.23188651}	&
	(580.5627)	&
	2013	&
	Ram and Bernath \cite{Ram2013}
	\\[1ex]
&%
	\bahiso{138}{1}	&
	3.3824544\(^{a(1)}\)	&
	2.23188\(^c\)	&
	(580.56260)	&
	1993	&
	Walker {\it et al} \cite{Walker1993}
	\\[1ex]
&%
	\bahiso{138}{1}	&
	3.382547\(^{a(1)}\)	&
	2.23185\(^c\)		&
	(569.64)	&
	1989		&
	Bernard {\it et al} \cite{Bernard1989}
	\\[1ex]
&%
	\bahiso{138}{1}	&
	3.382477	&
	2.2318987	&
	(580.597)	&
	1988	&
	Magg, Birk and Jones \cite{Magg1988}
	\\[1ex]
&%
	\bahiso{138}{1}	&
	3.382263	&
	2.23194\(^c\)	&
	(580.5673) &
	1985	&
	Appelblad {\it et al} \cite{Appelblad1985}
	\\[1ex]
&%
	\bahiso{138}{1}	&
	3.38285	&
	2.23175	&
	(580.53)	&
	1978	&
	Huber and Herzberg \cite{Huber1979}
	\\[1ex]
&%
	\bahiso{138}{1}	&
	3.38285	&
	2.23175\(^c\)	&
	(580.53)	&
	1966	&
	Kopp, Kronekvist and Guntsch \cite{Kopp1966-2}
	\\[1ex]
&%
	\bahiso{138}{2}	&
	\emph{1.7072}	&
	\emph{2.2303}\(^d\)	&
	(\emph{413.05})&
	1966	&
	Kopp and Wirhed \cite{Kopp1966-1}
	\\[1ex]
&%
	\bahiso{138}{1}	&
	3.3825\(^{a(1)}\)	&
	2.2319\(^c\)	&
	--	&
	1935	&
	Koontz and Watson \cite{Koontz1935}
\\[2ex]
H\(^2\Delta\) &
	\bahiso{138}{1}	&
	3.125\(^{a(3)}\)	&
	2.295\(^d\)	&
	9698.98	&
	2018	&
	Moore and Lane \cite{Moore2018}, theoretical	\rule{0pt}{2ex}
	\\[1ex]
&%
	\bahiso{138}{1}	&
	3.2174225	&
	2.288404	&
	9242.8	&
	1992	&
	Allouche {\it et al} \cite{Allouche1992} citing Bernard {\it et al} \cite{Bernard1989}
	\\[1ex]
&%
	\bahiso{138}{1}	&
	\textbf{3.217423}\(^{a(1)}\)&
	\textbf{2.28840}\(^c\)	&
	9243.13	&
	1989	&
	Bernard {\it et al} \cite{Bernard1989}
	\\[1ex]
H\(^2\Delta_{5/2}\) &
	\bahiso{138}{1}	&
	3.14998\(^{a(1)}\)	&
	2.31277\(^c\)	&
	8888.644\(^{a(1)}\) 	& 
	1987	&
	Fabre {\it et al} \cite{Fabre1987}
	\\[1ex]
&%
	\bahiso{138}{1}	&
	2.97	&
	--	&
	10609	&
	1978	&
	Huber and Herzberg \cite{Huber1979}
	\\[2ex]
A\(^2\Pi\) &
	\bahiso{138}{1}	&
	3.280\(^{a(3)}\)	&
	2.279\(^d\)	&
	10076.45	&
	2018	&
	Moore and Lane \cite{Moore2018}
	\\[1ex]
&%
	\bahiso{138}{1}	&
	\textbf{3.3004}\(^b\)	&
	\textbf{2.25945}\(^c\)	&
	9698.64	&
	1966	&
	Kopp, Kronekvist and Guntsch \cite{Kopp1966-2}
	\\[1ex]
&%
	\bahiso{138}{1}	&
	3.25965	&
	2.273533	&
	9727.2	&
	1992	&
	Allouche {\it et al} \cite{Allouche1992} citing Bernard {\it et al} \cite{Bernard1989}
	\\[1ex]
&%
	\bahiso{138}{1}	&
	3.25965\(^{a(1)}\)	&
	2.27353\(^c\)	&
	9728.67	&
	1989	&
	Bernard {\it et al} \cite{Bernard1989}
	\\[1ex]
&%
	\bahiso{138}{1}	&
	3.300\(^b\)	&
	2.249	&	
	9698.64\(^b\)	&
	1978	&
	Huber and Herzberg \cite{Huber1979}
	\\[2ex]
B\(^2\Sigma^+\)	&
	\bahiso{138}{1}	&
	3.270\(^{a(3)}\)	&
	2.291	&
	11112.61	&
	2018	&
	Moore and Lane \cite{Moore2018}
	\\[1ex]
&%
	\bahiso{138}{1}	&
	\textbf{3.268795}	&
	\textbf{2.27035}\(^c\)		&
	11092.5926	&
	1985		&
	Appelblad {\it et al} \cite{Appelblad1985}
	\\[1ex]
&%
	\bahiso{138}{1}	&
	3.21525	&
	2.321905	&
	10992.3	&
	1992	&
	Allouche {\it et al} \cite{Allouche1992} citing Bernard {\it et al} \cite{Bernard1989}
	\\[1ex]
&%
	\bahiso{138}{1}	&
	3.215253\(^{a(1)}\)	&
	2.28918\(^c\)		&
	10993.31	&
	1989	&
	Bernard {\it et al} \cite{Bernard1989}
	\\[1ex]
&%
	\bahiso{138}{1}	&
	3.164 (3.266)\(^f\)	&
	2.308	&
	11092.44	&
	1978	&
	Huber and Herzberg \cite{Huber1979} citing Veseth \cite{Veseth1973}
	\\[1ex]
&%
	\bahiso{138}{2}	&
	\emph{1.609 (1.636)}\(^f\)	&
	\emph{2.298}	&
	\emph{11089.62}	&
	1978	&
	Huber and Herzberg \cite{Huber1979} citing Veseth \cite{Veseth1973}
	\\[1ex]	
&%
	\bahiso{138}{2}	&
	\emph{1.6355}	&
	\emph{2.2787}\(^d\)	&
	\emph{11089.60}	&
	1966	&
	Kopp and Wirhed \cite{Kopp1966-1}
	\\[1ex]
&%
	\bahiso{138}{1}	&
	3.2682\(^{a(2)}\)	&
	2.2706\(^c\)	&
	--	&
	1935	&
	Koontz and Watson \cite{Koontz1935}
\\[1ex]
&%
	\bahiso{138}{1}	&
	3.232\(^{e}\)	&
	--	&
	--	&
	1933	&
	Watson \cite{Watson1933}
\\[2ex]
E\(^2\Pi\)	&
	\bahiso{138}{1}	&
	3.522\(^a\)	&
	2.190	&
	14871.07	&
	2018	&
	Moore and Lane \cite{Moore2018}
	\\[1ex]
&%
	\bahiso{138}{1}	&
	\textbf{3.520609}	&
	\textbf{2.187651}	&
	14830.1578	&
	2013	&
	Ram and Bernath \cite{Ram2013}
	\\[1ex]
&%
	\bahiso{138}{1}	&
	3.48510\(^e\)	&
	2.19877\(^e\) &
	14859.889\(^e\)	&
	1987	&
	Fabre {\it et al} \cite{Fabre1987}
	\\[1ex]
&%
	\bahiso{138}{1}	&
	3.523	&
	2.187	&
	14830	&
	1978	&
	Huber and Herzberg \cite{Huber1979}
	\\[1ex]
\hline
\noalign{\vskip 1.6mm}
\multicolumn{7}{p{15.5cm}}{\(^a\) Extrapolated from vibrational level values. The number in parentheses is the highest (\(v+\tfrac{1}{2}\)) term in the power series.}\\
\multicolumn{7}{l}{\(^b\) Determined from the average value of \(\Omega\) states.}\\
\multicolumn{7}{l}{\(^c\) \(r_e\) determined from corresponding \(B_e\).}\\
\multicolumn{7}{l}{\(^d\) \(r_e\) determined from a spline interpolation of the potential-energy curve.}\\
\multicolumn{7}{l}{\(^e\) Only \(v=0\) is measured, so the reported values refer to that vibrational level.}\\
\multicolumn{7}{l}{\(^f\) Following deperturbation analysis (uncorrected value in brackets).}\\
\end{tabular}
\end{table*}

\begin{table*}
\centering
\footnotesize
\caption{Decay pathways from the B\(^2\Sigma_{1/2}^+, v=0, N=0\) excited state of BaH. $\mathcal{A}$ is the Einstein $A$ coefficient $A_{v'v''}$ for each transition and $\mathcal{R}$atio is the value of $\mathcal{R}_{v'v''}$ from Eq. (\ref{eq:exp_VBR}).
Schemes included are
(i) $T_0$ shifted but \(\Delta r_e\) uncorrected from \textit{ab initio} potentials used in Moore18 \cite{Moore2018} and (ii, iii) \(d\)-complex potentials shifted to match experimental values of the \(\Delta r_e\) value compared to the X\(^2\Sigma^+\) state.
Scheme ii uses the data of Appelblad {\it et al} \cite{Appelblad1985} for the B\(^2\Sigma^+\) state while scheme
iii uses the data of Bernard {\it et al} \cite{Bernard1989}.  The calculated emission VTR is based on Eq. (\ref{eq:VBR}) and the calculated lifetime refers to the lowest rovibrational level. The recommended values, based on the present measurements of the branching ratio, are shown in bold.}\label{tab:bah_Bstate_re}
\vspace{2mm}
\begin{tabular}{|l|lr|lr|lr|}
\hline
\rule{0pt}{2.5ex}Final rovibronic	&	 \multicolumn{2}{|c|}{(i) $T_0$ corrected only}	&	 \multicolumn{2}{|c|}{(ii) Appelblad {\it et al}}	&	 \multicolumn{2}{|c|}{\bf{(iii) Bernard {\it et al}}}\\
state	&
\multicolumn{1}{c}{\(\mathcal{A}\) / s\(^{-1}\)}	&
\multicolumn{1}{c|}{\(\mathcal{R}\)atio}	&
\multicolumn{1}{c}{\(\mathcal{A}\) / s\(^{-1}\)}	&
\multicolumn{1}{c|}{\(\mathcal{R}\)atio}	&
\multicolumn{1}{c}{\bf{\(\mathcal{A}\) / s\(^{-1}\)}}	&
\multicolumn{1}{c|}{\bf{\(\mathcal{R}\)atio}}
\rule{0pt}{2.5ex}\\\hline
X\(^2\Sigma^+_{1/2}, v=0, N=1\)	&
	7.85\(\times\)10\(^6\)	&	97.255\%	&
	8.05\(\times\)10\(^6\)	&	98.553\%	&
	\textbf{7.76\(\times\)10\(^6\)}	&	\bf{96.648\%}
\rule{0pt}{2.5ex}\\
X\(^2\Sigma^+_{1/2}, v=1, N=1\)	&
	2.14\(\times\)10\(^5\)	&	2.656\%	&
	1.11\(\times\)10\(^5\)	&	1.363\%	&
	\bf{2.62\(\times\)10\(^5\)}	&	\bf{3.256\%}
\rule{0pt}{2.5ex}\\
X\(^2\Sigma^+_{1/2}, v=2, N=1\)	&
	1.51\(\times\)10\(^3\)	&	0.009\%	&
	3.51\(\times\)10\(^2\)	&	0.004\%	&
	\textbf{1.25\(\times\)10\(^3\)}	&	\bf{0.016\%}
\rule{0pt}{2.5ex}\\
H\(^2\Delta_{3/2}, v=0, J=\tfrac{3}{2}\)	&
	2.39\(\times\)10\(^3\)	&	0.030\%	&
	2.40\(\times\)10\(^3\)	&	0.029\%	&
	\textbf{2.40\(\times\)10\(^3\)}	&	\bf{0.030\%}
\rule{0pt}{2.5ex}\\
A\(^2\Pi_{1/2}, v=0, J=\tfrac{1}{2}\)	&
	1.55\(\times\)10\(^3\)	&	0.019\%	&
	1.58\(\times\)10\(^3\)	&	0.019\%	&
	\textbf{1.53\(\times\)10\(^3\)}	&	\bf{0.019\%}
\rule{0pt}{2.5ex}\\
A\(^2\Pi_{1/2}, v=0, J=\tfrac{3}{2}\)	&
	1.33\(\times\)10\(^3\)	&	0.016\%	&
	1.34\(\times\)10\(^3\)	&	0.016\%	&
	\textbf{1.34\(\times\)10\(^3\)}	&	\bf{0.017\%}
\rule{0pt}{2.5ex}\\
A\(^2\Pi_{3/2}, v=0, J=\tfrac{3}{2}\)	&
	1.18\(\times\)10\(^3\)	&	0.015\%	&
	1.20\(\times\)10\(^3\)	&	0.015\%	&
	\textbf{1.17\(\times\)10\(^3\)}	&	\bf{0.015\%}
\rule{0pt}{2.5ex}\\[.5ex]
\hline	Calculated VTR &
\multicolumn{2}{c|}{0.038}&
\multicolumn{2}{c|}{0.019}&
\multicolumn{2}{c|}{\bf{0.047}}
\rule{0pt}{2.5ex}\\[.5ex]
\hline	Calculated lifetime &
\multicolumn{2}{c|}{\(\tau = \)123.9~ns}&
\multicolumn{2}{c|}{\(\tau = \)122.5~ns}&
\multicolumn{2}{c|}{\bf{\(\tau = \)124.3~ns}}
\rule{0pt}{2.5ex}\\
\hline
\end{tabular}
\end{table*}

\subsection{Potential energy curves}\label{PECurves}

For initial simulations of the branching ratios, the potentials calculated by Moore18 \cite{Moore2018} were used.  These {\it ab initio} calculations of the potential-energy curves were performed at a post Hartree-Fock level using a parallel version of the
\texttt{MOLPRO} \cite{Werner2010,Werner2012} (version 2010.1) suite of quantum chemistry codes. The  aug-cc-pCV$Q$Z (ACV$Q$Z) basis set \cite{Li2013} was used on the barium atom to describe the 5$s$5$p$6$s$ electrons, and the equivalent aug-cc-pV$Q$Z basis set was used for hydrogen~\cite{Dunning1989}. An effective core potential \cite{Lim2006} was used to describe the lowest 46 core electrons of the barium atom. The active space at long-range corresponded to the occupied valence orbitals plus the excited 6$p$5$d$ and the lowest Rydberg 7$s$ orbital on barium. Once the Hartree-Fock wave function had been found, the electron correlation was determined using both the state-averaged complete active space self-consistent field~\cite{Siegbahn1980} and the multireference configuration interaction (MRCI) \cite{Knowles1988} methods for static and dynamic correlation, respectively.  Higher levels of correlation were approximated using the Davidson correction \cite{Davidson1974}. The MRCI wave functions were then used to calculate transition dipole moments (TDMs) and spin-orbit coupling matrix elements using the \texttt{MOLPRO} code \cite{Berning2000}.  Further details on the potentials involved in the present paper, namely the ground X$^2\Sigma^{+}_{1/2}$ and excited H$^2\Delta_{{3/2}}$, A$^2\Pi_{{1/2},{3/2}}$ and B$^2\Sigma^{+}_{1/2}$ states shown in Fig. \ref{spin_rotation}(a), can be found in Moore18 \cite{Moore2018}.  As in Moore18, the traditional Hund's case (a) electronic label is used for those potentials calculated without consideration of spin-orbit coupling such as B$^2\Sigma^{+}$, while the form
A$^2\Pi_{1/2}$ is used for the final states where $\Omega$, the projection of the total electronic angular momentum on the internuclear axis, is a good quantum number.

To calculate the rovibrational energy levels in each state, the radial Schr{\"o}dinger equation for each {\it ab initio} potential is solved using the program \texttt{DUO} \cite{Yurchenko2016}.  To include the most significant interactions between the states of interest, the treatment includes the electronic states X\(^2\Sigma^+\), H\(^2\Delta\), A\(^2\Pi\), B\(^2\Sigma^+\) and E\(^2\Pi\) and the relevant spin-orbit matrix elements \cite{Berning2000}.
The relevant rovibrational levels are shown in Fig. \ref{spin_rotation}(b) for the B$^2\Sigma_{1/2}^+$ $v=0$ $\leftarrow$ X$^2\Sigma_{1/2}^+$ $v=1$ repump transition.  While \texttt{MOLPRO} represents all calculations in the \(C_{2v}\) point group symmetry, \duo\ handles  \(C_{\infty v}\) symmetry states, so appropriate transformations \cite{Patrascu2014} are required to prepare \texttt{MOLPRO} output data for input into \duo as described in Moore18. The spectroscopic values of $T_e$ (the energy of the potential minimum) and $r_e$ for each electronic state of interest are presented in Table \ref{exp_ref}.

\subsection{Determining the branching ratios}\label{CAtransition}

The program \texttt{DUO} \cite{Yurchenko2016} was also used to determine the decay pathways and branching ratios from the {\it ab initio} potentials and assorted calculated matrix elements \cite{Patrascu2014}.
The lifetime of each rovibronic state is calculated using the MRCI TDMs \cite{Berning2000}. The TDMs involved in both the A$^2\Pi$ $\rightarrow$ X$^2\Sigma^+$ and B$^2\Sigma^+$ $\rightarrow$ X$^2\Sigma^+$ transitions are of similar magnitude and hence the lifetimes of these two states are likely to be comparable.  Also strong is the TDM connecting the A$^2\Pi$ and H$^2\Delta$ states, but B$^2\Sigma^+$ $\rightarrow$ A$^2\Pi$ is considerably weaker than the other three.  These minor TDMs are important because they are responsible for the main radiative loss pathways for the A$^2\Pi_{1/2}$ -- X$^2\Sigma^+_{1/2}$ and B$^2\Sigma^+_{1/2}$ -- X$^2\Sigma^+_{1/2}$ cooling cycles \cite{Moore2018}. The presence of spin-orbit mixing introduces a new and significant decay pathway B\(^2\Sigma^+_{1/2} \rightarrow  \) H$^2\Delta_{3/2}$ that competes with the Laporte-allowed decay channels of the B$^2\Sigma^+_{1/2}$ state, as detailed in Table~\ref{tab:bah_Bstate_re}.

Besides the TDM, the quantities that affect the lifetimes are the FC factors, and the difference in equilibrium bond length $\Delta r_e=r_e^{'}-r_e^{''}$ between the upper and lower states is especially important for determining these.  The decay rate associated with an electronic transition can be expressed in terms of the Einstein $A$ coefficient, defined within the \texttt{DUO} code \cite{Yurchenko2016} as
\begin{multline} \label{eq:A_coeff}
A_{v'v''} =  \frac{16 \pi^{3}}{3 \epsilon_{0}h} (2J'' + 1) \tilde{\nu}_{v'v''}^{3}  \\
\sum_{n}^{all} \left[ (-1)^{\Omega''} \small \begin{pmatrix}
J''      & 1   & J' \\
\Omega'' & n & \Omega'
\end{pmatrix}
{ \mel{ v'}{\mu_{n} (r)}{ v''}} \right]^{2}
\end{multline}
where $\mel{ v'}{\mu_{n} (r)}{ v''}$ is the vibrationally averaged transition dipole moment, the matrix represents a 3$j$ symbol, and the summation is over all components of the transition dipole.
Due to the vibrational averaging, any error in the calculated values of $T_e$ and $r_e$ reduces the accuracy of the calculated decay rates and the excited-state lifetimes. To minimize these errors, the potentials were shifted to match experimental measurements of these quantities (see Appendix) whenever possible.

\subsubsection*{$T_e$ corrections}

The first correction applied to the excited states is to match the transition energies to experimental data.  In Moore18, the correction to $T_0$ was made by replacing the calculated transition frequencies for each decay channel with the spectroscopic values from the literature after analyzing the raw {\it ab initio} data in \texttt{DUO}.  This corrects the ${\tilde{\nu}_{v'v''}}^{3}$ term in Eq. (\ref{eq:A_coeff}) for the Einstein $A$ coefficient.

In this paper, we iteratively adjusted the energy of the potential minimum in \texttt{DUO} until the calculated  \(T_0\) separation was within 0.01\wn\ of the spectroscopic values.  To ensure the correct energetic displacement between the  X\(^2\Sigma^+\) and B\(^2\Sigma^+\) states, the experimental work of Appelblad \textit{et al} \cite{Appelblad1985} was initially used as a reference for the separation between the \(T_0\) energies in each state. Similarly, spectroscopic data from Kopp {\it et al} \cite{Kopp1966-2} were used for the A$^2\Pi$ state, and those from Ram and Bernath \cite{Ram2013} were used for the  E$^2\Pi$ state.  For H$^2\Delta$, the experimental data from Bernard {\it et al} \cite{Bernard1989} for the spin-orbit splitting were used in combination with the spectroscopic H$^2\Delta_{5/2}$ $\nu^{\prime}$~=~0, $J$ = 5/2 energy determined in Ref. \cite{Fabre1987}.

The {\it ab initio} $T_e$ for the B$^2\Sigma^+_{1/2}$ state calculated by Moore18 is in remarkable agreement with the majority of spectroscopic studies and just 120 cm$^{-1}$ higher than the relative outlier by Bernard {\it et al}. The situation is different for the A$^2\Pi_{1/2}$  state, first as there are larger discrepancies between the spectroscopic values, and second since the theoretical value is too high, by as much as 380 cm$^{-1}$. As a consequence, the calculated energy difference $\Delta E =$ $E$(A$^2\Pi_{1/2} -$ B$^2\Sigma^+_{1/2}$) is in error by 20-25\%. The final energy shifts applied are -459.46 \wn (H$^2\Delta$), -385.06 \wn (A$^2\Pi$) and -99.32 \wn (B$^2\Sigma^+$). The resulting calculated spectroscopic constants are shown in Table~\ref{tab:To_shift}.  In particular, the energy shift to the A$^2\Pi$ state significantly improves the calculated spin-rotation constant in the B$^2\Sigma^+$ state, as outlined in the Appendix.

The accuracy of the {\it ab initio} results for the B$^2\Sigma^+$ state is particularly striking. The calculated value for the lowest vibrational energy separation ($\Delta G_{10}$) in the B$^2\Sigma^+$ state (1057.16 cm$^{-1}$) agrees within 0.09\% of the observed value (1058.04 cm$^{-1}$). This is a surprisingly good match to experiment for the relatively small quadruple-zeta basis set, while the performance for the other electronic states is more in line with expectations.  This change in methodology has the effect of slightly, but occasionally significantly, modifying the calculated decay rates (Table \ref{tab:bah_Bstate_re}) compared to those reported in Moore18.  These rates can be expressed as raw Einstein $A$ coefficients for each decay channel or as the ratio of that channel to the total overall decay rate, $\mathcal{R}_{v'v''}$ in Eq.~(\ref{eq:exp_VBR}), a  useful parameter when assessing the viability of laser cooling.  No modification for a primary decay channel exceeds 0.07\%.  However, for the B$^2\Sigma^+_{1/2}$ state there is a significant reduction in the decay to H$^2\Delta$ (from 0.054 to 0.030\%) and to $v =$ 2 of the ground state (from 0.014 to 0.009\%).  These revisions affect the fine details of the cooling efficiency (compare the ``$T_0$ corrected only" column in Table~\ref{tab:bah_Bstate_re} with Table 6 from Moore18) but do not alter the overall findings from Moore18.

\begin{table}
\centering
\footnotesize
\caption{Vibronic state parameters as determined from the \duo\ analysis of the {\it ab initio} results, following energy shifts to the {\it ab initio} MRCI+Q potentials calculated with the ACV$Q$Z basis set on Ba.  The simulation includes the lowest five electronic states (X$^2\Sigma^+$, H$^2\Delta$ , A$^2\Pi$, B$^2\Sigma^+$ and E$^2\Pi$) and all the relevant spin-orbit and ladder matrix elements. These shifted potentials match spectroscopic $T_0$ values for each state.  $A_v$ is the spin-orbit splitting, $\gamma _v$ is the spin-rotation constant and all values are in \wn. The X$^2\Sigma^+$ zero-point energy of 580.5673 cm$^{-1}$ from Appelblad {\it et al} \cite{Appelblad1985} is used for the B$^2\Sigma^+$ state. The spectroscopic values from Appelblad {\it et al}  are shown in bold. }
\label{tab:To_shift}
\vspace{2mm}
\begin{tabular}{|ll|rrrrr|}
\hline
\rule{0pt}{2.9ex}State	&	$v$	\rule{0pt}{2ex}	&
	$T_v$	&	$A_v$	&	$B_v$	&	10$^4D_v$	&
	$\gamma_v$	
	\\[1ex]\hline
\rule{0pt}{3.0ex}\Xst	\rule{0pt}{2ex}
&	0	&	    0.00	&	--	
		&  3.3271	&	  1.1359	&	  0.2226		\\
&	1	&	 1125.72	&	--	
		&  3.2622	&	  1.1147	&	  0.2175		\\
&	2	&	 2227.62	&	--	
		&  3.1976	&	  1.1302	&	  0.2124		\\
&	3	&	 3299.75	&	--	
		&  3.1332	&	  1.1107	&	  0.2073		\\[1ex]
\Ast	\rule{0pt}{2ex}
&	0	&	 9664.34	&	  483.63	
		&  3.2393	&	  1.2188	&	--		\\
&	1	&	10742.01	&	  483.77	
		&  3.1654	&	  1.2888	&	--		\\
&	2	&	11789.43	&	  481.49	
		&  3.0641	&	 -0.2486	&	--		\\
&	3	&	12809.63	&	  488.40	
		&  3.0050	&	  0.9440	&	--		\\[1ex]
\Bst	\rule{0pt}{2ex}
&	0	&	11052.60	&	--	
		&  3.2307	&	  1.1324	&	 -4.9039	\\
&		&  \bf{11052.61}\(^a\)  &		
		&  \bf{3.2334}	& \bf{1.1570} &	\bf{-4.7539} \\
&	1	&	12109.76	&	--	
		&  3.1644	&	  1.1261	&	 -4.7924	\\
&		&  \bf{12110.64}    &		
		&  \bf{3.1627}	& \bf{1.1541} &	\bf{-4.6343} \\
&	2	&	13135.59	&	--	
		&  3.0956	&	  1.1401	&	 -4.6923	\\
&		&  \bf{13137.94} &		
		&  \bf{3.0919}	& \bf{1.1522} &	\bf{-4.5178} \\
&	3	&	14132.49	&	--	
		&  3.0264	&	  1.1650	&	 -4.5856	\\
&		&  \bf{14134.65} &		
		&  \bf{3.0211}	& \bf{1.1582} &	\bf{-4.3897} \\[.5ex]
\hline
\noalign{\vskip 1.4mm}
\multicolumn{7}{l}{\(^a\)Using X$^2\Sigma^+$ zero-point energy from Appelblad {\it et al} \cite{Appelblad1985}.
\rule{0pt}{1.5ex}}\\
\end{tabular}
\end{table}

\subsubsection*{$r_e$ corrections}

\begin{figure}
\center
\includegraphics[width=90mm]{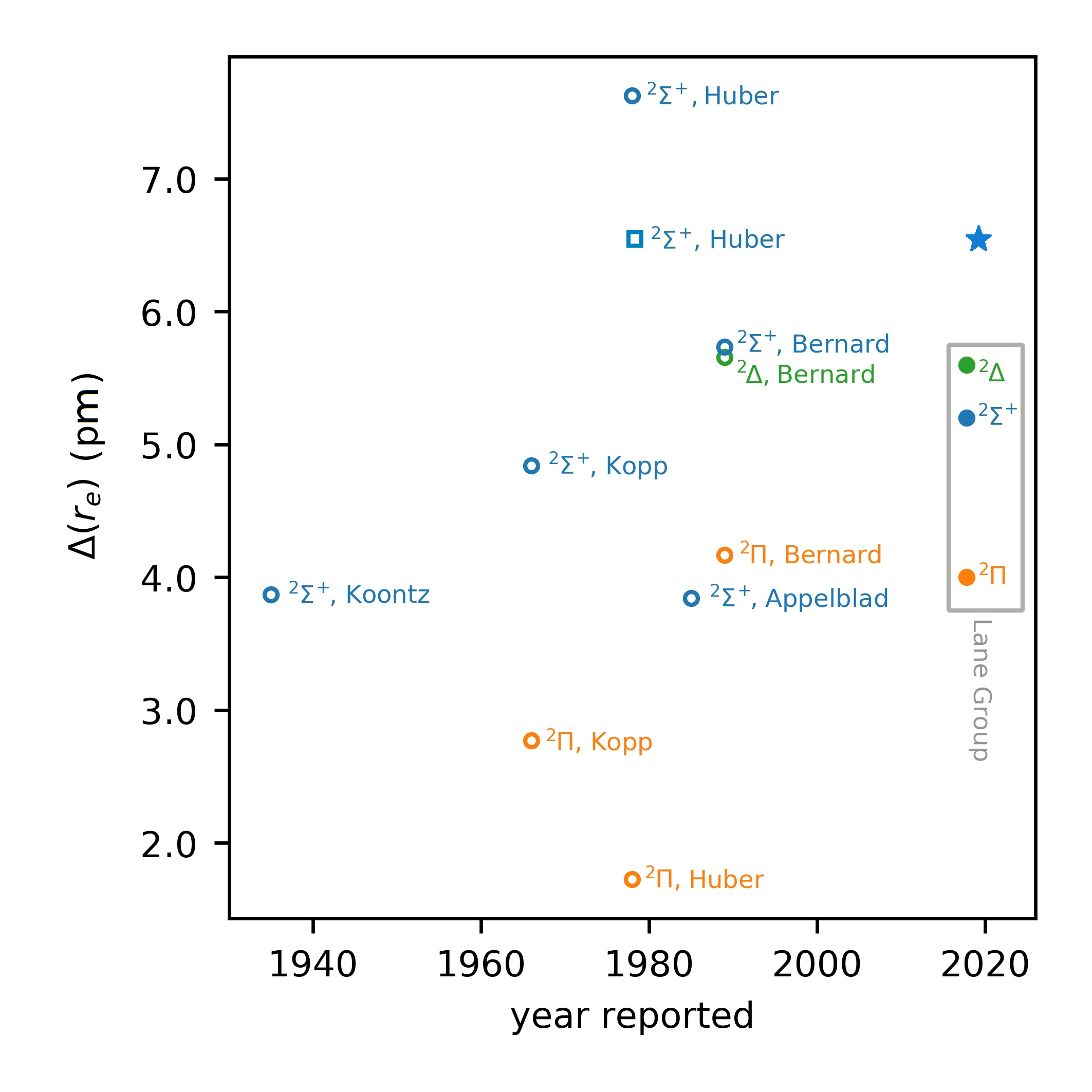}
\caption{A timeline for the observed values of $\Delta r_e$, for the lowest excited states H$^2\Delta$ (green), A$^2\Pi$ (orange), and B$^2\Sigma^+$ (blue) of BaH.  The unfilled (blue) square corresponds to $\Delta r_e$ for the B state of BaD reported by Veseth \cite{Veseth1973} (and published in Huber and Herzberg \cite{Huber1979}).  Also marked are the {\it ab initio} values of Moore18 \cite{Moore2018} calculated with MRCI wave functions using the  ACV$Q$Z basis set. The recommended B$^2\Sigma^+$ value, based on the branching-ratio measurements and {\it ab initio} calculations presented here, is indicated by a star.}
\label{fig:Delta_re}
\end{figure}

Most of the branching-ratio measurements here involve the  B -- X  transition and so corrections to these two potentials are important.
The consensus ground-state bond length has been determined to 0.01-pm accuracy as 2.2319 \AA, while the last four experiments yield 2.23188 \AA\ within 0.005 pm.  This value is used to determine $\Delta r_e$ for each experiment that presented $r_e$ values for the $5d$-complex excited states.  These differences are tabulated in Table \ref{tab:bah_delta_re} and plotted in Figure \ref{fig:Delta_re}.  The theoretical values from Moore18 are also included for reference.

As discussed earlier, there are two recent spectroscopic studies of the B$^2\Sigma^+$ state. In the first,
the $r_{e}$ is reported by Appelblad {\it et al} \cite{Appelblad1985} following their spectroscopic analysis and is consistent with the earlier $r_e$ measurements \cite{Watson1933,Koontz1935} at the 0.1-pm level. The second study by Bernard {\it et al} \cite{Bernard1989} reports only the rotational constants for the first two vibrational levels.  Therefore, the higher-order fitting terms (see Appendix) are neglected, and the extrapolation becomes
\begin{equation} \label{eq:exp_re}
B_e = \tfrac{3}{2} B_0 - \tfrac{1}{2} B_1.
\end{equation}
Equation (\ref{eq:exp_re}) was used in Moore18 to determine the measured  $r_e$ values of the excited states where the spectroscopic data was limited. However, Veseth \cite{Veseth1973} has argued that a simple extrapolation of the observed constants is inadequate in the case of a strongly interacting system like the A$^2\Pi$ and B$^2\Sigma^+$ states in BaH, and produced revised rotational constants based on a more sophisticated model. Using the data from Koontz and Watson \cite{Koontz1935}, for example, this analysis suggests $B_e=3.164$ cm$^{-1}$ as quoted in Huber and Herzberg \cite{Huber1979}, smaller (thus indicating a longer bond length r$_{e}$) than the originally published value. This suggests that a linear fit of the Bernard data using Eq. (\ref{eq:exp_re}) may underestimate the  B$^2\Sigma^+_{1/2}$ bond length.

The $r_e$ of Appelblad {\it et al} \cite{Appelblad1985} is nearly 1.9 pm shorter than the value determined using the experimental data of Bernard {\it et al} \cite{Bernard1989} with Eq. (\ref{eq:exp_re}).  However, the earliest work on the B$^2\Sigma^+_{1/2}$ spin-rotation constant \cite{Watson1933,Koontz1935} reports a relatively large and negative value (i.e. the level with the higher value of $J$ lies at a lower energy) $\gamma_0 = -4.84$ cm$^{-1}$ (the subscript refers to the $v=0$ level) while Bernard {\it et al} \cite{Bernard1989} suggests a much smaller and positive value of $\gamma_0$ = 0.46149 cm$^{-1}$, reversing the ordering of the levels. The contemporaneous value from Appelblad {\it et al} is more precise than the older data \cite{Watson1933,Koontz1935} but is clearly in agreement, $\gamma_0 = -4.7538$ cm$^{-1}$. It is also in agreement with the present {\it ab initio} results.  Therefore, based solely on spectroscopy, the most reliable experimental data comes from Appelblad {\it et al} \cite{Appelblad1985} and consequently was the basis for the experimental corrections used in Moore18.

\begin{table}
\center
\caption{$\Delta r_e$, the difference between the equilibrium bond lengths in the excited and ground states, from a variety of spectroscopic studies.  Theoretical work of Moore18 \cite{Moore2018} determined this value by spline interpolation of  MRCI+Q {\it ab initio} points  using the ACV$Q$Z basis set.}
\label{tab:bah_delta_re}
\center
\footnotesize
\begin{tabular}{|llll|}
\noalign{\vskip 2mm}
\hline
\rule{0pt}{3ex}State &
	\multicolumn{1}{c}{\(\Delta(r_e)\) / pm} &
	Year &
	\multicolumn{1}{c|}{Reference}
	\rule{0pt}{2ex}\\
\hline
\rule{0pt}{4ex}H\(^2\Delta\)	&
	+5.6	&
	2018	&
	Moore and Lane \cite{Moore2018}	\rule{0pt}{2.5ex}
	\\[1ex]
&%
	+5.655	&
	1989	&
	Bernard {\it et al} \cite{Bernard1989}
	\\[2ex]
A\(^2\Pi\)	&
	+4.0	&
	2018	&
	Moore and Lane \cite{Moore2018}
	\\[1ex]
&%
	+4.168	&
	1989	&
	Bernard {\it et al} \cite{Bernard1989}	
	\\[1ex]
&%
	+1.725	&
	1978	&
	Huber and Herzberg \cite{Huber1979}	
	\\[1ex]
&%
	+2.77	&
	1966	&
	Kopp, Kronekvist and Guntsch \cite{Kopp1966-2}
	\\[2ex]
B\(^2\Sigma^+\)	&
	+5.2	&
	2018	&
	Moore and Lane \cite{Moore2018}
	\\[1ex]
&%
	+9.003	&
	1992	&
	Allouche {\it et al} \cite{Allouche1992}
	\\[1ex]		
&%
	+5.733	&
	1989	&
	Bernard {\it et al} \cite{Bernard1989}
	\\[1ex]
&%
	+3.841	&
	1985	&
	Appelblad {\it et al} \cite{Appelblad1985}
	\\[1ex]
&%
	+7.63	&
	1978	&
	Huber and Herzberg \cite{Huber1979}
	\\[1ex]
&%
	+6.63	&
	1973	&
	Veseth\(^a\) \cite{Veseth1973}
	\\[1ex]
&%
	+4.84	&
	1966	&
	Kopp and Wirhed \cite{Kopp1966-1}
	\\[1ex]
&%
	+3.87	&
	1935	&
	Koontz and Watson \cite{Koontz1935}
	\\[1ex]
\hline
\noalign{\vskip 1.6mm}
\multicolumn{4}{l}{\(^a\) BaD, following deperturbation analysis.}\\
\end{tabular}
\end{table}

As B$^2\Sigma^+_{1/2}$ levels can decay to the lower A$^2\Pi$ and H$^2\Delta$ states, each with two spin-orbit components (no decay can take place to H$^2\Delta_{5/2}$), it is also prudent to consider shifting the internuclear separation of these potential minima to the spectroscopic values to ensure that the calculated rates are as accurate as possible. For both components of the  H$^2\Delta$ state, the only recent experimental value is from Bernard {\it et al} \cite{Bernard1989}.
This spectroscopic $\Delta (r_e)$ is 0.055 pm different for the theoretical value, so in all simulations the H$^2\Delta_{3/2}$ potential is shifted by 0.05 pm. The equivalent shift for the A$^2\Pi_{1/2}$ state was the same as that used in the spin-rotation calculation, namely the raw {\it ab initio} value, as discussed below.  All the tabulated theoretical values are obtained using these transformations.

\begin{figure}[!ht]
\center
\includegraphics[width=80mm]{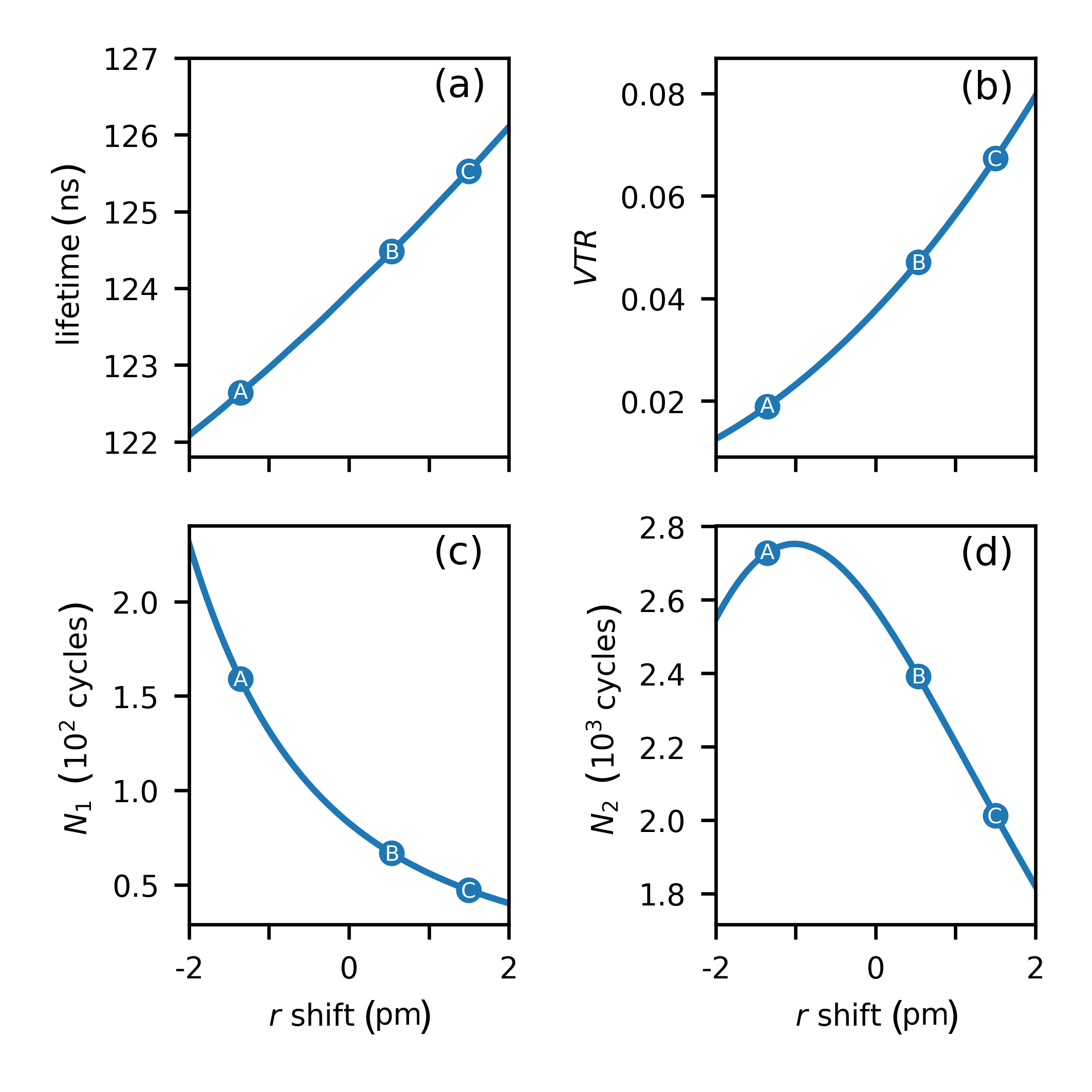}
\caption{Fundamental laser cooling parameters for the B$^2\Sigma^+$ $\leftarrow$ X$^2\Sigma^+$ transition computed with MRCI wave functions using the ACV$Q$Z basis set.  The T$_0$ values have been corrected to the experimental values as described in the text.
The panels present the effect of small shifts in the $r_e$ value of the B$^2\Sigma^+_{1/2}$ state on the lifetime (a), the vibronic transition ratio (VTR, Eq. (\ref{eq:VBR})) for decay to X$^2\Sigma^+_{1/2}$ $v =$ 1 vs $v =$ 0 (b), and the number of one-color cooling cycles $N_1$ (c) and two-color cooling cycles $N_2$ (d). The $r_e$ shifts are relative to the {\it ab initio} $\Delta r_e^{B-X}$.
The shifts in the {\it ab initio} bond length required to match the experimental difference $\Delta r_e^{B-X}$ are shown with filled dots: Appelblad {\it et al} data \cite{Appelblad1985} (A), Bernard {\it et al} data \cite{Bernard1989} (B), and our branching-ratio data leading to the proposed longer B$^2\Sigma^+$ bond length (C).
}
\label{fig:Shift_effects}
\end{figure}

The effect of adjusting $r_e$ for the B$^2\Sigma^+_{1/2}$ potential on its lifetime is shown in Fig.~\ref{fig:Shift_effects}(a).  There is a smooth but significant sensitivity to change in the B$^2\Sigma^+$ state $r_e$ (with respect to the {\it ab initio} result) even over a range of just $\pm$ 2 pm ($\pm$  0.85 \% of the actual bond length).  A much stronger relative effect can be seen in the emission VTR \cite{Zak2017} of Fig.~\ref{fig:Shift_effects}(b),
\begin{equation} \label{eq:VBR}
\mathrm{VTR} = \frac{q_{01}}{q_{00}} =  \sum_{J ^{\prime \prime}}{ \left(\frac{S_{0J',1J''}^{\vphantom{q}}}{S_{0J',0J''}^{\vphantom{q}}}\right) }
\end{equation}
where $q_{0i}$ is the FC factor for the transition from B$^2\Sigma^+$ $v^{\prime} = 0, J ^{\prime} $ = 1/2 to the X$^2\Sigma^+$ $v^{\prime \prime} = i$ vibrational level ($i=$ 0, 1), and $S_{0J',iJ''}^{\vphantom{q}}$ is the line strength factor as in Eq. (\ref{eqn:linestrength}) that is summed over the decays to both $J^{\prime \prime}$ = 1/2 and 3/2 of ($-$) parity.
The experimental bond-length shift $\Delta r_e^{B-X}$ from Appelblad {\it et al} \cite{Appelblad1985} corresponds to the dot marked A in all panels of Fig.~\ref{fig:Shift_effects}.  In Moore18, in an effort to minimize the effect of errors in $\Delta r_e^{B-X}$, the calculated B$^2\Sigma^+$ state is shifted to this spectroscopic value prior to determination of the decay channels. However, the agreement with the measured emission VTR of 0.092(20) is much worse than for the uncorrected potentials (Table~\ref{tab:bah_Bstate_re}).

To resolve this problem, the first step was to verify whether the chosen spectroscopic data is consistent with other experimental studies and {\it ab initio} calculations.  Comparison with the {\it ab initio} potentials from Moore18 (Table  \ref{tab:bah_delta_re}) reveals a better agreement between the theoretical results and the spectroscopic $\Delta r_e$ values reported by Bernard {\it et al} \cite{Bernard1989} for all three 5$d$-complex states, assuming that Eq.(\ref{eq:exp_re}) is an acceptable extrapolation of the recorded B$_v$ constants.
Therefore, an additional analysis was performed by shifting all the potentials to be consistent with the measurements in Bernard {\it et al}.
The experimental $T_e$ is $\sim100$ cm$^{-1}$ different from Appelblad {\it et al} used in Moore18.  This small energy difference has a negligible effect on the decay rates, and this adjustment to the energy was not applied.  Far more important was the $\Delta r_e$ for B$^2\Sigma^+_{1/2}$.
By using the Bernard {\it et al} value of $\Delta r_e^{B-X}$ = +5.733 pm (dots marked B in Fig. \ref{fig:Shift_effects}) the agreement with the experimental branching ratio has significantly improved.
The theoretical A$^2\Pi_{1/2}$ potential was a close enough match to the Bernard {\it et al} value of +4.168 pm not to warrant any shift to the {\it ab initio} value (Table \ref{tab:bah_Bstate_re}).
The present simulation is a superior match to experiments:  in particular, the calculated lifetime of the B$^2\Sigma^+_{1/2}$ state is now 124.3 ns, in excellent agreement with the lifetime of the $J$~=~11/2 level measured in Ref.~\cite{Berg1997} and an improvement on the uncorrected potentials.
The shift in $r_e$ significantly increases the decay to X$^2\Sigma^+_{1/2}$ $v=2$, to an extent that it becomes comparable to radiative decay to A$^2\Pi_{1/2}$ $v=0$, $J=3/2$ but is still below the decay to H$^2\Delta_{3/2}$.

\subsection{Refining the potentials}\label{Compare-cool}

In addition to $\Delta r_e$, the shape of the potential-energy function $V(r)$ affects the accuracy of the vibronic transition moments through changes to the vibrational wave functions.  The presently calculated aug-cc-pCV$Q$Z $V(r)$ for B$^2\Sigma^+_{1/2}$ is of a very high quality for a quadruple-zeta potential.
Describing the barium atomic orbitals using the aug-cc-pCV$n$Z basis sets ($n$ = $Q$, 5) taken to the complete basis set (CBS) limit \cite{Moore2016} produces a high-quality ground state potential X$^2\Sigma^+$.  The calculated $r_e$, for example, lies within 0.03 pm of the experimental value of Ram and Bernath \cite{Ram2013}.  Furthermore, by fitting this potential to spectroscopic data using \texttt{DPotFit} \cite{LeRoy2017a}, including the correct dispersion behavior of the potential at extended bond lengths \cite{Derevianko2010}, a very accurate potential can be generated from the short range to the atomic asymptote.  The {\it ab initio} points were first fitted to a 13-parameter MLR potential \cite{LeRoy2006,LeRoy2009} using \texttt{betaFIT} (version 2.1) \cite{LeRoy2017b} and then combined with X$^2\Sigma^+$ infrared experiments \cite{Walker1993} and B$^2\Sigma^+$ $\rightarrow$ X$^2\Sigma^+$ emission data \cite{Watson1933, Koontz1935, Kopp1966-1} for processing with \texttt{DPotFit}.  No measurements from Bernard {\it et al} \cite{Bernard1989} were used in this fitting process. As the data set contains both BaH and BaD spectroscopic information, the isotopic Born-Oppenheimer breakdown corrections \cite{LeRoy2002} could also be determined. The final rovibrational levels for the lowest three vibrational quantum states can be reproduced with an accuracy better than 0.005 cm$^{-1}$. Replacing the present ACV$Q$Z ground state with the MLR potential from Ref. \cite{Moore2016} should improve the accuracy of the calculated transitions.

\begin{table}
\centering
\caption{Effect of adjusting the {\it ab initio} B$^2\Sigma^+$ and X$^2\Sigma^+$ potentials on laser cooling.  The FC factors are calculated using the program \texttt{DUO} \cite{Yurchenko2016}. The top row corresponds to the excited B$^2\Sigma^+_{1/2}$ state  represented by the ACV$Q$Z potentials from Moore18, where the bond-length difference $\Delta r_e$ is set at the experimental values derived from Bernard {\it et al} \cite{Bernard1989}. The ground state used is the ACV$n$Z/CBS MLR potential of Ref. \cite {Moore2016}.  The bottom row adds a further extension to the excited-state bond length so that it differs by + 1.5(1) pm from the {\it ab initio} result.}

\label{tab:bah_further_adjustments}
{\footnotesize
\begin{tabular}{|l|c|c|c|c|}
\noalign{\vskip 3mm}
\hline
\rule{0pt}{2.5ex}	& \multicolumn{4}{|c|}{B\(^2\Sigma^+_{1/2}\) $-$ X\(^2\Sigma^+_{1/2}\) FC factors }    		\\
\hline
\rule{0pt}{2.5ex}  Adjustment & (00) \hspace{0.2mm} & \hspace{0.3mm} (01)	& (02) $\times10^{3}$	& (03) $\times10^{6}$ 	\\[0.2ex]
\hline
\rule{0pt}{3ex}	
Bernard $\Delta r_e^{B-X}$	&	 0.937 &	0.062	& 0.65	&	 1.4	\rule{0pt}{0.2ex}	\\
\rule{0pt}{3ex}		
+ 1.5(1) pm	&	 0.922(4) &	0.077(4)	&  1.09(9) 	&	 2.5(2)	\rule{0pt}{0.1ex}	\\
	
\hline
\noalign{\vskip 1.6mm}
\end{tabular}}
\end{table}

\begin{figure}[!ht]
\center
\includegraphics[width=90mm]{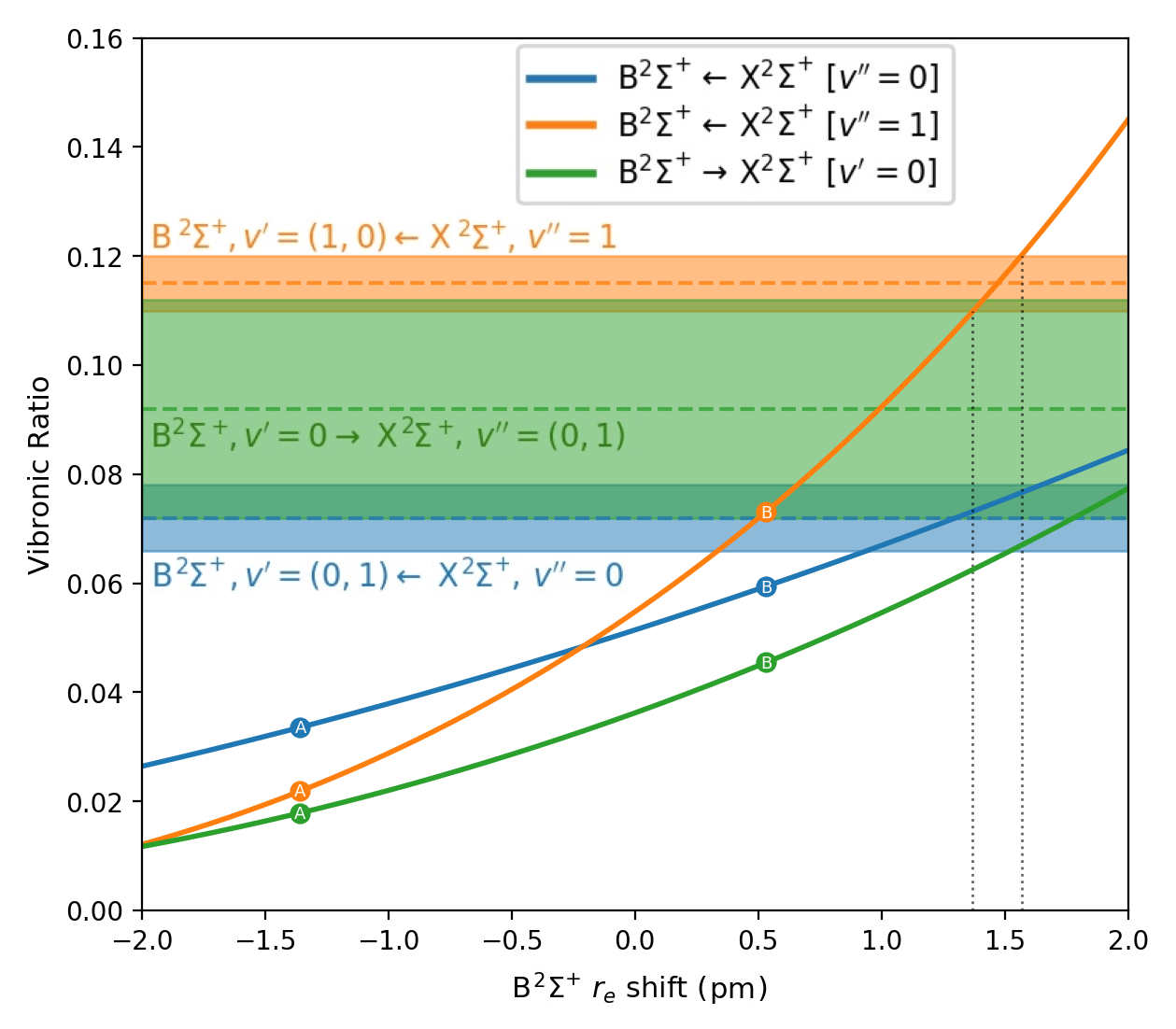}
\caption{Comparison between our B$^2\Sigma^+$ $\leftarrow$ X$^2\Sigma^+$ VTR (vibronic transition ratio) measurements (horizontal dashed lines, with uncertainties represented by the horizontal bars) and the predicted ratios based on {\it ab initio} results (solid lines). The ground state is the MLR potential based on ACV$n$Z/CBS MLR calculations from Ref. \cite{Moore2016} and the excited state is based on a ACV$Q$Z potential. The difference in equilibrium bond lengths between the two states is varied between $-2$ and $+2$ pm from the {\it ab initio} result $\Delta r_e^{B-X}$ = 5.2 pm.  The blue data (lowest horizontal bar) corresponds to  $q_{10}$/$q_{00}$ and the orange data (highest horizontal bar) corresponds to $q_{01}$/$q_{11}$.  The green data corresponds to $q_{01}$/$q_{00}$.  Also marked are the shifts in the {\it ab initio} bond length required to match the experimental difference $\Delta r_e^{B-X}$ with the upper state value of Appelblad {\it et al} \cite{Appelblad1985} (dots A) while dots B are set at the value of Bernard {\it et al}  \cite{Bernard1989}.
The black dotted vertical lines represents the range of bond-length corrections where the calculated and measured VTRs agree.}
\label{fig:Abs_shifts}
\end{figure}

The absorption data consists of VTR measurements to the B$^2\Sigma^+$ $v=0$ and $1$ levels from $v=0$ or $1$ in the ground X$^2\Sigma^+$ state.  Specifically, the $N^{\prime}=0$ $J^{\prime} =\frac{1}{2}$ (+) $\leftarrow$ $N^{\prime \prime}=1$ $J^{\prime \prime} =\frac{1}{2}$ ($-$) absorption lines ($Q_{12}$) were measured for each vibrational band, $q_{v'v''}$ determined for each transition and the ratios $q_{10}$/$q_{00}$ and $q_{01}$/$q_{11}$ found (Table \ref{tab:bah_comparisons}).
These experimental FC factors can be compared directly with the theoretical $S_{v'v''}$ ($Q_{12}$ lines only so the reference to $J$ can be dropped) line strength factors.  The theoretical branching ratios can be plotted as a function of the deviation in $\Delta r_e^{B-X}$ from the {\it ab initio} value (+5.2 pm) as in Fig. \ref{fig:Shift_effects} and then compared with measured VTRs as shown in Fig. \ref{fig:Abs_shifts}. The plot clearly demonstrates the highly sensitive dependence of $q_{01}$/$q_{11}$ in particular on the bond length and that the two experimental $\Delta r_e^{B-X}$ values \cite{Appelblad1985,Bernard1989} (dots marked A and B in Fig. \ref{fig:Abs_shifts}) predict ratios that are lower than observed.
We find that any discrepancies between the calculated and experimental branching ratios can be resolved by shifting the potential minimum of the B$^2\Sigma^+$ state to a longer bond length.  To match experiment within error bars, a bond-length difference
$\Delta r_e^{B-X}$ = 6.5 - 7.0 pm is necessary, requiring a B$^2\Sigma^+_{1/2}$ state bond length 0.8 - 1.3 pm greater than derived from Bernard {\it et al} but still somewhat shorter (see Fig. \ref{fig:Delta_re}) than the values reported in Huber and Herzberg \cite{Huber1979} ($\Delta r_e^{B-X}$ = 7.6 pm) and Allouche {\it et al} \cite{Allouche1992} (9.0 pm).  This is, however, consistent with the value for BaD  derived by Veseth \cite{Veseth1973} (by conducting a detailed re-analysis of the data from Kopp and Wirhed \cite{Kopp1966-1}) that is also quoted in Huber and Herzberg \cite{Huber1979}.

\begin{table}
\centering
\caption{Radiative decay pathways from the lowest rovibronic states of B$^2\Sigma^+$ and A\(^2\Pi_{1/2}\) for both $v^\prime$ = 0 and 1.  The A\(^2\Pi_{1/2}\) $v^\prime$ = 0 level is involved in the principal laser cooling while B$^2\Sigma^+$ $v^\prime$ = 0 aids in repumping.
The alternative repumping scheme involving A\(^2\Pi_{1/2}\) $v^\prime$ = 1 is compromised by a large decay to $v''=2$.
$\mathcal{A}$ is the Einstein $A$ coefficient for each transition and $\mathcal{R}$atio is the value of $\mathcal{R}_{v'v''}$.  The excited states are represented by the ACV$Q$Z potentials from Moore18, and $\Delta r_e$ is set at the experimental value derived from Bernard {\it et al} \cite{Bernard1989} (H$^2\Delta$ state), the raw {\it ab initio} value (A$^2\Pi$ state), and the BaD value reported by Veseth~\cite{Veseth1973} (B$^2\Sigma^+$ state).  The ground state is the ACV$n$Z/CBS MLR potential of Ref. \cite {Moore2016}. The decay from B$^2\Sigma^+$ $v^\prime$ = 0 to X$^2\Sigma^+$ $v^{\prime \prime}$ = 0 and 1 is further decomposed into the $Q_{12}$ and $P_{1}$ sub-branches. The radiative lifetime of each excited level is indicated.
}

\label{tab:bah_decay_branching}
{\footnotesize
\begin{tabular}{|l|llrr|}
\noalign{\vskip 3mm}
\hline
\rule{0pt}{2.5ex}	& \multicolumn{4}{|c|}{Decay pathways}		\\
\rule{0pt}{2.5ex}   & Final state  & \(v''\)	& \(\mathcal{A}\) / s\(^{-1}\)\hspace{1mm}	& $\mathcal{R}$atio / \%\hspace{1.2mm}	\\[0.2ex]
	
\hline
\rule{0pt}{3ex}A\(^2\Pi_{1/2}\)  		
	&	X\(^2\Sigma^+_{1/2}\) 	($N=1$)  	&	0	& $7.23\times10^6$	&	98.772	\rule{0pt}{2.5ex}	\\
\(v' =\) 0
	&	X\(^2\Sigma^+_{1/2}\) 	&	1	& $8.94\times10^4$	&	 1.221	\rule{0pt}{2.5ex}	\\
	&	X\(^2\Sigma^+_{1/2}\) 	&	2	& $1.50\times10^2$	&	 0.002	\rule{0pt}{2.5ex}	\\
	&	H\(^2\Delta_{3/2}\) ($J= \frac{3}{2}$) 	&	0	& $3.24\times10^2$	&	 0.004	\rule{0pt}{2.5ex}	\\
	&	\textbf{Lifetime} (ns)	&		&  	&	 \bf{136.5}  \rule{0pt}{2.5ex}	\\
	
\hline
\rule{0pt}{3ex}B\(^2\Sigma^+_{1/2}\)		
	&	X\(^2\Sigma^+_{1/2}\) 	($N=1$) &	0	&  $7.61\times10^6$	&	 95.312	\rule{0pt}{2.5ex}	\\
\(v' =\) 0	 &  $\vphantom{QQ}  $ \(Q_{12}\) 	&		&  {\it $4.13\times10^6$}	&	 {\it 51.704}	\rule{0pt}{2.5ex}	\\
	 &	 $\vphantom{QQ}  $ \(P_{1}\) 	&		&  {\it $3.48\times10^6$}	&	 {\it 43.609}	\rule{0pt}{2.5ex}	\\
	 &	X\(^2\Sigma^+_{1/2}\) 	&	1	&  $3.64\times10^5$	&	 4.564	\rule{0pt}{2.5ex}	\\
	 &   $\vphantom{QQ}  $ \(Q_{12}\) 	&		&  {\it $2.88\times10^5$}	&	 {\it 3.611}	\rule{0pt}{2.5ex}	\\
	 &	 $\vphantom{QQ}  $ \(P_{1}\) 	&		&  {\it $7.62\times10^4$}	&	 {\it 0.953}	\rule{0pt}{2.5ex}	\\
	&	X\(^2\Sigma^+_{1/2}\) 	&	2	&  $3.44\times10^3$	&	 0.043		\rule{0pt}{2.5ex}	\\
	&	A\(^2\Pi_{1/2}\) 	($J= \frac{1}{2}$) 	 &	0	&  $1.48\times10^3$	&	 0.019	\rule{0pt}{2.5ex}	\\
	&	A\(^2\Pi_{1/2}\) 	($J= \frac{3}{2}$)	 &	0	&  $1.37\times10^3$	&	 0.017	\rule{0pt}{2.5ex}	\\
	&	A\(^2\Pi_{3/2}\) 						 &	0	&  $1.14\times10^3$	&	 0.014	\rule{0pt}{2.5ex}	\\
	&	H\(^2\Delta_{3/2}\) 					 &  0	&  $2.39\times10^3$	&	 0.030	\rule{0pt}{2.5ex}	\\
	&	\textbf{Lifetime} (ns)	&		&  	&	 \bf{125.1}  \rule{0pt}{2.5ex}	\\
	
\hline
\rule{0pt}{3ex}A\(^2\Pi_{1/2}\)  		
	&	X\(^2\Sigma^+_{1/2}\) 	($N=1$)  	&	0	& $6.91\times10^5$	&	9.400	\rule{0pt}{2.5ex}	\\
\(v' =\) 1
	&	X\(^2\Sigma^+_{1/2}\) 	&	1	& $6.47\times10^6$	&	 87.952	\rule{0pt}{2.5ex}	\\
	&	X\(^2\Sigma^+_{1/2}\) 	&	2	& $1.93\times10^5$	&	 2.633	\rule{0pt}{2.5ex}	\\
	&	X\(^2\Sigma^+_{1/2}\) 	&	3	& $5.60\times10^2$	&	 0.008	\rule{0pt}{2.5ex}	\\		
	&	A\(^2\Pi_{1/2}\) 	(both $J$)	 		&	0	&  $1.45\times10^2$	&	 0.002	\rule{0pt}{2.5ex}	\\	
	&	H\(^2\Delta_{3/2}\) ($J= \frac{3}{2}$) 	&	1	& $3.62\times10^2$	&	 0.005	\rule{0pt}{2.5ex}	\\	
	&	\textbf{Lifetime} (ns)	&		&  	&	 \bf{135.9}  \rule{0pt}{2.5ex}	\\
		
\hline
\rule{0pt}{3ex}B\(^2\Sigma^+_{1/2}\)		
	&	X\(^2\Sigma^+_{1/2}\) 	($N=1$) 	&	0	&  $8.81\times10^5$	&	 11.595	\rule{0pt}{2.5ex}	\\
\(v' =\) 1
	 &	X\(^2\Sigma^+_{1/2}\) 	&	1	&  $6.16\times10^6$	&	81.081	\rule{0pt}{2.5ex}	\\
	&	X\(^2\Sigma^+_{1/2}\) 	&	2	&  $5.39\times10^5$	&	 7.100	\rule{0pt}{2.5ex}	\\
	&	X\(^2\Sigma^+_{1/2}\) 	&	3	&  $1.00\times10^4$	&	 0.132	\rule{0pt}{2.5ex}	\\
	&	B\(^2\Sigma^+_{1/2}\) 			  	&	0	&  $9.41\times10^1$	&	 0.001	\rule{0pt}{2.5ex}	\\		
	&	A\(^2\Pi_{1/2}\) 		(both $J$)	&	0	&  $5.88\times10^2$	&	 0.008	\rule{0pt}{2.5ex}	\\
	&	A\(^2\Pi_{1/2}\) 			 		&	1	&  $2.63\times10^3$	&	 0.035	\rule{0pt}{2.5ex}	\\
	&	A\(^2\Pi_{3/2}\) 	($J= \frac{3}{2}$)		&	0	&  $4.07\times10^2$	&	 0.005	\rule{0pt}{2.5ex}	\\
	&	A\(^2\Pi_{3/2}\) 							&	1	&  $1.01\times10^3$	&	 0.013	\rule{0pt}{2.5ex}	\\	
	&	H\(^2\Delta_{3/2}\) 					 	&   1	&  $2.24\times10^3$	&	 0.029	\rule{0pt}{2.5ex}	\\
	&	\textbf{Lifetime} (ns)	&		&  	&	 \bf{131.7}  \rule{0pt}{2.5ex}	\\

\hline
\noalign{\vskip 1.6mm}
\end{tabular}}
\end{table}

The $q_{10}/q_{00}$ branching ratio was also measured for the A$^2\Pi_{3/2}$ $\leftarrow$ X$^2\Sigma^+$ transition via absorption.
When computing this VTR, the A$^2\Pi_{3/2}$ {\it ab initio} ACV$Q$Z potential without $r_e$ adjustment (as it is almost identical to Bernard {\it et al}) is used to simulate the measurement.  The calculated $q_{10}/q_{00}$ ratio is in excellent agreement with the experimental result (Table \ref{tab:bah_comparisons}).  Unlike for the B$^2\Sigma^+$ state, however, this result is not a good match to the re-analysis on A$^2\Pi_{3/2}$ by Veseth (r$_{e} =$ 2.259 \AA
~\cite{Veseth1973} for BaD, or 1.0 pm longer than the  Huber and Herzberg \cite{Huber1979} value for BaH indicated in Fig. \ref{fig:Delta_re}).

Using the superior ground-state potential does not improve the agreement between the calculated and measured emission VTR $q_{01}/q_{00}$ (last line in Table \ref{tab:bah_comparisons}), lowering the theoretical B$^2\Sigma^+$ $-$ X$^2\Sigma^+$ value to 0.045 compared with 0.047 with the ground ACV$Q$Z potential (Table  \ref{tab:bah_Bstate_re}).  Any remaining discrepancies between the calculated and experimental branching ratios can be resolved by shifting the potential minimum of the B$^2\Sigma^+$ state to a longer bond length as with the absorption data.
The arithmetic mean for the bond increase determined from the absorption and emission data is +1.5(1) pm.  The improved agreement with the VTR measurements is presented in Table \ref{tab:bah_comparisons}.

The poorer agreement observed in the B -- X emission calculations is removed when only the ratios of $Q_{12}$ lines are compared (as in the absorption simulations). Adopting the adjustment to the Veseth~\cite{Veseth1973} (BaD) $\Delta r_e^{B-X}$, the ratio of emission linestrengths $S_{01}$/$S_{00}$ is 0.069 for the $Q_{12}$ lines while this falls to 0.022 for the $P_{1}$ lines.  The former value is in very good agreement with experiment and with the calculated FC factors (Table \ref{tab:bah_further_adjustments}), perhaps indicating that the calculated $P_{1}$  line for the $v^{\prime}=0$ -- $v^{\prime \prime}=1$ transition is too weak.

\section{Laser cooling implications for BaH}\label{Compare-cool}

The superior modeling of the branching-ratio data using the extended B$^2\Sigma^+$ bond length indicates that to ensure an improvement in the simulation of the cooling process by correcting for experimental data the optimal experimental correction would rely on the BaD value reported by Veseth~\cite{Veseth1973} over those of Appelblad {\it et al} and  Bernard {\it et al}.  Adopting this value, the decay rates were computed for the lowest rovibrational levels in all three 5$d$-complex states.  The updated branching ratios confirm the conclusion of Moore18 that the A$^2\Pi$ $\leftarrow$ X$^2\Sigma^+$ transition is advantageous for Doppler cooling despite the longer wavelength of the transition, as shown in Table \ref{tab:bah_decay_branching}.
Crucially, the A$^2\Pi$ -- X$^2\Sigma^+$ cooling has a single loss channel involving another electronic state (H$^2\Delta$) while the  B$^2\Sigma^+$ -- X$^2\Sigma^+$ transition has four decay routes, all significantly stronger (Table~\ref{tab:bah_decay_branching}).
One notable feature of the present results is the relative strength of decay to X$^2\Sigma^+$ $v^{\prime \prime} =2$.

The molecular parameters relevant to Doppler cooling are collected in Table \ref{tab:bah_parameters}.
The lifetime of  H$^2\Delta_{3/2}$ is determined by spin-orbit mixing with the A$^2\Pi$ state and the strength of the A$^2\Pi$ -- X$^2\Sigma^+$ TDM.  The former is significantly reduced over the original calculation in Moore18, so the lifetime of the H$^2\Delta_{3/2}$ state has almost doubled to 9.5 $\mu$s.  The other lifetimes are nearly unchanged, while the associated slight increase in  B$^2\Sigma^+_{1/2}$ lifetime (125.1 ns,  still in excellent agreement with experiment) and reduction in the maximum number of the B$^2\Sigma^+$ -- X$^2\Sigma^+$ cooling cycles ($N_n$) are documented in both Fig. \ref{fig:Shift_effects} (dots C) and Table \ref{tab:bah_parameters}.
The number of cycles $N_n$ that $n$ light fields can support to laser cool a fraction $\mathcal{F}$ of the molecules depends on the branching ratios $\mathcal{R}_{v^{\prime} v^{\prime \prime}}$ \cite{Lane2015,Moore2018}.
Setting $\mathcal{F}$ to 0.1 (90\% loss in molecular-beam intensity), the number of cycles supported by each transition can be determined for one-, two-, and three-color cooling (the third transition involves either excitation out of the H$^2\Delta_{3/2}$ state \cite{Moore2018} or from X$^2\Sigma^+$ $v^{\prime \prime} =$ 2 depending on which loss channel is greater).  The change in $\Delta r_e$ has had a particularly large effect on the number of one-color cooling cycles that can be supported on the A --X and B -- X (Fig.~\ref{fig:Shift_effects}(c)) cooling transitions.  In both cases, this number has been profoundly lowered.
However, using a second laser to repump $v^{\prime \prime} =$ 1 creates a nearly closed cycle (Fig.~\ref{fig:Shift_effects}(d)), in particular for the A$^2\Pi$ $\leftarrow$ X$^2\Sigma^+$ transition (over 99.99\% of the population recycled).

Shifting the bond length in  A$^2\Pi_{1/2}$ from the experimental value of Kopp {\it et al.} \cite{Kopp1966-2} to Bernard {\it et al.} \cite{Bernard1989} does have a somewhat detrimental effect on the efficiency of the A$^2\Pi$ -- X$^2\Sigma^+$ cooling cycle. It is still very effective, but the number of cycles possible with two cooling lasers $N_2=3.5\times10^4$ is rather smaller than reported in Moore18 \cite{Moore2018}. Naturally, the increased B$^2\Sigma^+$ state $r_{e}$ also reduces the value of $N_2$ ($2.3\times10^3$) from the previous calculation, but the relative reduction is somewhat smaller.  This is partly because the increased B$^2\Sigma^+$ bond length ($\Delta r_e^{B-X}$ = 6.7 pm) does indeed raise the decay to X$^2\Sigma^+$ $v^{\prime \prime} =$ 2 but this is ameliorated by the reduced decay to H$^2\Delta_{3/2}$ (the largest decay channel) which the present revised calculations predict. However, using a third laser to repump the B$^2\Sigma^+_{1/2}$ $v=2$ level will only yield a marginal improvement ($N_3=3.2\times10^3$) while it more than triples the corresponding number of A$^2\Pi$ -- X$^2\Sigma^+$ cycles.

\begin{table}
\centering
\caption{Revised calculated properties of the proposed laser cooling transitions in BaH molecules.
The differences in equilibrium bond lengths, $\Delta r_e$, are the recommended values, set at the experimental values derived from Bernard {\it et al} \cite{Bernard1989} for the H\(^2\Delta_{3/2}\) state, the {\it ab initio} value for A\(^2\Pi_{1/2}\), and both Bernard {\it et al} and Veseth (italics) \cite{Veseth1973} for B\(^2\Sigma^+_{1/2}\), the latter
closely matching the longer excited-state bond length proposed in this paper.
Here $N_i$ is the number of cycles supported by $i$ lasers before the population falls to 10\%,
$T_D$ and $v_D$ are the Doppler temperature and velocities,
$v_c$ is the capture velocity, and
the maximum deceleration is $a_{\mathrm{max}} = \hbar k A/(2M) = v_r A/2$.}

\label{tab:bah_parameters}
{\footnotesize
\begin{tabular}{|l|rrr|}
\noalign{\vskip 5mm}
\hline
Molecular 		\rule{0pt}{2.5ex}	& \multicolumn{3}{|c|}{State}		\\
\hspace{3mm}data	&	B\(^2\Sigma^+_{1/2}\)	\hspace{2.5mm}& \hspace{4.3mm} A\(^2\Pi_{1/2}\)	&	\hspace{3.5mm} H\(^2\Delta_{3/2}\)	\hspace{0.5mm} \\
\hline
\(\lambda\)/nm	
	&	905.3 	&	1060.8	&	1110
\rule{0pt}{2.5ex}	\\
\(\Delta r_{e}\)/pm
	&	+5.7 {\it (+6.7)} 	&	+4.0	&	+5.6
\rule{0pt}{2.5ex}	\\
\(\tau\)/ns$^a$
	&	124.3 {\it (125.1)} 	&	136.5	&	9532
\rule{0pt}{2.5ex}	\\
\(N_1\)	
	&	70 {\it (47)} 	&	186	 &	-
\rule{0pt}{2.5ex}	\\
\(N_2\) ($\times10^{3}$)
	&	2.3 {\it (1.9)}	&	35.3	&	-
\rule{0pt}{2.5ex}	\\
\(N_3\)	($\times10^{3}$)
	&	3.2 {\it (2.9)}	&	111	&	-
\rule{0pt}{2.5ex}	\\
\(T_D\)/\(\mu\)K	
	&	30.7 {\it (30.4)}	&	27.9	&	0.4	
\rule{0pt}{2.5ex}	\\
\(v_c\)/cm s\(^{-1}\)	
	&	116 {\it (115)}	&	124	&	1.8	
\rule{0pt}{2.5ex}	\\
\(v_D\)/cm s\(^{-1}\)	
	&	4.3 {\it (4.2)}	&	4.1	&	0.5	
\rule{0pt}{2.5ex}	\\
\(a_{\mathrm{max}}\)/ms\(^{-2}\)	
	&	12.7{\it (6)}~$\times10^{3}$	&	$9.9\times10^{3}$	&	-	
\rule{0pt}{2.5ex}	\\[1ex]
\hline
\noalign{\vskip 1.5mm}
\multicolumn{3}{l}{\textsuperscript{a} For the lowest rovibronic level.}\\
\end{tabular}}
\end{table}

Repumping both $v=0$ and $1$ levels in the X$^2\Sigma^+$ state via the lowest vibronic level of the A$^2\Pi_{1/2}$ state is the preferred option to keep radiative losses to a minimum. Practically, to take advantage of maximum radiation pressure forces, implementations of laser cooling often avoid transitions that share the same excited level. Unfortunately, pumping the $v^{\prime}=1$ level in the A$^2\Pi_{1/2}$ (or B$^2\Sigma^+_{1/2}$) state will significantly increase the decay rate to $v^{\prime \prime}=2$ and will even open decay to $v^{\prime \prime}=3$ \cite{Moore2018}.
Consequently, a combination of transitions involving  A$^2\Pi$ -- X$^2\Sigma^+$  (0 -- 0) and  B$^2\Sigma^+$ -- X$^2\Sigma^+$  (0 -- 1) is the best compromise cooling scheme by virtue of having the lowest additional losses (approximately 0.0014\% losses per cycle), although the increased excited-state loss from B$^2\Sigma^+_{1/2}$ means that the two-color efficiency is reduced to $N_{2} = 2.86\times10^4$ cycles.

One additional loss channel to consider is possible vibrational decay via infrared emission within the X$^2\Sigma^+$ ground state, particularly from the $v=1$, $N=1$ level as it participates in the two-color cooling cycle.  The relatively large vibrational spacing in hydrides results in a significant Einstein $A$ coefficient over rival ultracold diatomics such as ionic fluorides or alkali metal dimers.
The upper $J^{\prime}=3/2$ ($-$) level can decay via three radiative transitions while there are two pathways for the lower $J^{\prime}=1/2$. In both cases, the decay to $v''=0,N^{\prime \prime}=2$, $J^{\prime \prime}>$ $J^{\prime}$ leads to the largest loss rate.
The total decay rates for $J^{\prime}=3/2$ and $J^{\prime}=1/2$ are 79.2 and 79.3 s$^{-1}$, respectively, resulting in practically identical radiative lifetimes of 12.6 ms.  This vibrational loss can be mitigated by operating the repumping lasers well above saturation and thus ensuring that the molecules spend minimal time in the X$^2\Sigma^+$ $v=1$ state.  This radiative pathway could also allow us to populate the absolute ground rovibronic state $v=0$, $J=1/2$ ($+$) after cooling by pumping into $v=1$, loading the molecules into a conservative trap, and waiting a short time for them to decay.  This is a unique feature of diatomic hydrides (as compared to other laser cooling candidates) due to their larger vibrational spacing.

\section{Conclusions}
\label{sec:Outlook}
The branching ratios of diagonal molecular transitions are extremely sensitive to the relative bond lengths between the excited and ground states.  In this paper,
the vibrational branching ratios for the B$^2\Sigma^+$ $\leftarrow$ X$^2\Sigma^+$ transition in BaH molecules were measured via optical fluorescence and absorption, and calculated using {\it ab initio} quantum chemistry methods.  By shifting the excited-state potential
by just 0.5 pm ($0.25\%$ of the bond length), an improved agreement with former experiments is achieved, e.g., with spectroscopic $r_e$ values for the excited states from Bernard {\it et al} \cite{Bernard1989}.
Furthermore, our measured branching ratio indicates that the {\it ab initio} excited-state B$^2\Sigma^+$ potential should be shifted by +1.5(1)pm relative to the ground state, in agreement with the BaD result quoted by Veseth~\cite{Veseth1973}.
This bond-length correction is found to have implications for the closure of the B$^2\Sigma^+$ state with respect to laser cooling, confirming the A$^2\Pi_{1/2}$ excited state as a superior choice.  The sensitivity of branching ratios to small changes in $\Delta r_e$, and the substantial inconsistency in the estimates of this parameter present in the literature, show that care must be taken when identifying potential laser cooling candidates.

\begin{acknowledgments}
\section{Acknowledgments}
The authors would like to thank N. Dattani, I. Kozyryev, and D. J. Owens for useful discussions and S. Vazquez-Carson for experimental assistance.  TZ would like to thank ONR Grant No. N00014-17-1-2246 and AFOSR Grant No. FA9550-17-1-0441-DURIP.
KM and ICL thank the Leverhulme Trust (Research Grant No. RPG-2014-212) for financial support including the funding of a studentship for KM.  RLM gratefully acknowledges support by the NSF Integrative Graduate Education and Research Traineeship Grant No. DGE-1069240.
\end{acknowledgments}

\section{Appendix}

\subsection{Spin-rotation coupling constants}
\label{Bstate}

The spectroscopic values of $T_e$ and $r_e$ for each electronic state of interest are presented in Table \ref{exp_ref}.  Typically, spectroscopic studies quote $T_{00}$, the energy difference between the $v=0$ levels of the upper and lower electronic states, or an extracted value of  $T_e$.  By fitting the reported vibronic energies, $T_e$ can be determined and there is a broad consistency between these measurements over the multiple studies.

In general, the agreement between spectroscopic values is less satisfactory for $r_e$ than for $T_e$. Where \(r_e\) is not directly quoted in a reference, it may be calculated from \(B_e\), the equilibrium rotational constant, using
\begin{equation}
r_e = \left( \frac{h}{8\pi^2\mu c B_e} \right)^{1/2},
\end{equation}

\noindent
where \(\mu\) is the reduced mass in kg, \(c\) is the speed of light in cm~s\(^{-1}\), $h$  is Planck's constant in J$\cdot$s, and \(B_e\) is the rotational constant in cm\(^{-1}\). In cases where only \(B_v\) values are quoted in the reference, \(B_e\) is extrapolated as
\begin{equation}
B_v = B_e - \alpha_e (v+\tfrac{1}{2}) + \delta_e (v+\tfrac{1}{2})^2.
\end{equation}
For many spectroscopic studies of BaH the strongly diagonal nature of the transitions often means that only the lowest two vibrational levels are resolved. Consequently, the square term is neglected and
equation (\ref{eq:exp_re}) must be used to determine the measured  $B_e$ values. Thus the accuracy of the final value of $r_e$ determined via this expression is naturally limited.

While the computed rovibrational spacings are in excellent agreement with experiment \cite{Moore2018}, properties that are particularly sensitive to either $r_e$ or $T_e$ may need to be corrected.  An example is the spin-rotation structure calculated for each $^2\Sigma^+_{1/2}$ state.  The spin-rotation constant $\gamma$ is typically decomposed into first- and second-order terms. The first-order spin-rotational coupling is due to the magnetic fields created by the rotation of the electric charge distribution of the molecule (it depends mainly on the component of the spin-orbit operator that includes nuclear momenta), but this contribution is frequently small.  Therefore, the spin-rotation constant is predominantly a second-order correction \cite{Brown2003} to the energy levels of a $^2\Sigma^+$ state found by summing over $^2\Pi$ states,
\begin{equation}
\gamma = \sum_{n^2\Pi}^{\infty} \small \frac{ \langle BL^+\rangle\langle\hat{H}_{\mathrm{SO}}\rangle}{\huge \Delta E(n ^2\Pi - ^2\Sigma^+)},
\end{equation}
where $\mel{^2\Sigma^+,0,-\tfrac{1}{2}}{B L^{+} }{n ^2\Pi,-1,+\tfrac{1}{2}}\equiv\langle BL^+\rangle$ are ladder matrix elements, $\mel{n ^2\Pi,-1,+\tfrac{1}{2}}{\hat{H}_{\mathrm{SO}}}{^2\Sigma^+,0,-\tfrac{1}{2}}\equiv\langle\hat{H}_{\mathrm{SO}}\rangle$ are spin-orbit matrix elements, and $ \Delta E(n ^2\Pi - ^2\Sigma^+)$ is the energy separation between each $^2\Pi$ state in the summation and the $^2\Sigma^+$ state. Both required matrix elements are computed using the spin-orbit package within \texttt{MOLPRO}. However, the energy separation between states is crucial, especially when relatively small.  For the B$^2\Sigma^+_{1/2}$ state, the largest contributions are from A$^2\Pi_{1/2}$ and E$^2\Pi_{1/2}$, the former 1540 cm$^{-1}$ lower in energy and the latter more than twice this value above the B$^2\Sigma^+_{1/2}$ minimum. As the matrix elements have similar magnitudes (Table 5 in Moore18), the A$^2\Pi_{1/2}$ state has the largest influence on the observed spin-rotation splitting. Consequently, for a reliable calculation of these constants it is important to ensure that the energy separation between the theoretical curves matches the experimental value.

Using {\it ab initio} potentials without adjustment, the spin-rotation constant computed by \texttt{DUO} is $\gamma_0 = -5.6295$ cm$^{-1}$. The A$^2\Pi$ potential was then lowered by 385 cm$^{-1}$ to match the measured $T_0$ value from  Ref. \cite{Kopp1966-2} and the spin-rotation constant was recalculated (Table \ref{tab:To_shift}). This adjusted value $\gamma_0 = -4.9039$ cm$^{-1}$ is in much better agreement with the value from Appelblad {\it et al} but appears to contradict the small, positive value found by Bernard {\it et al}.


\end{document}